\documentclass[a4paper,aps,nofootinbib,superscriptaddress,12pt]{revtex4}
\usepackage{amsmath}         
\usepackage{amssymb}          
\usepackage[pdftex]{graphicx}	
\usepackage{slashed}
\usepackage{color}
\usepackage{ulem}
\usepackage{dsfont}
\usepackage[pdftex]{hyperref}

\newcommand{\ord}[1]{\mathcal{O}\left({#1}\right)}

\begin{document}

\title{Can SUSY relax LNV constraints coming from loop corrections to light neutrino masses on the low-scale Seesaw?}

\author{J. Jones-P\'erez}
\email[E-mail: ]{jones.j@pucp.edu.pe}
\author{O. Suarez-Navarro}
\email[E-mail: ]{osuarez@pucp.edu.pe}
\affiliation{
Secci\'on F\'isica, Departamento de Ciencias, Pontificia Universidad Cat\'olica del Per\'u, Apartado 1761, Lima, Peru
}

\begin{abstract}
Heavy neutrinos from the Type-I Seesaw model can have a large mixing with active states, motivating their search at collider experiments. However, loop corrections to light neutrino masses constrain the heavy neutrinos to appear in pseudo-Dirac pairs, leading to a potential suppression of lepton number violating parameters.

In this work we perform a detailed review of a proposal to relax constraints on lepton number violation by adding supersymmetry. We define the conditions necessary to maximise the SUSY screening effect, with the objective of allowing a larger mass splitting between low-scale heavy neutrino masses. We find that the sole addition of SUSY does not guarantee a screening, and that favourable cases have some degree of fine-tuning.
\end{abstract}

\maketitle

\section{Introduction}

The Type I Seesaw~\cite{Minkowski:1977sc,GellMann:1980vs,Yanagida:1979as,Mohapatra:1979ia,Schechter:1980gr} is very likely the most studied extension of the Standard Model (SM) explaining neutrino masses. One of its key predictions is the existence of heavy neutrinos $N_h$, although their number, mass scale, and coupling strength, remain free parameters. This has motivated their search by several experiments (see reviews~\cite{Atre:2009rg,Deppisch:2015qwa,Abdullahi:2022jlv}), with unfortunately null signals to date.

It is well known that, in its most basic realisation, the Seesaw is actually very hard to test. The $N_h$ interact via their mixing with the active flavour states ($\nu_e,\,\nu_\mu,\,\nu_\tau$), and the typical expectation is that the square of this mixing will be proportional to the ratio between light and heavy neutrino masses, out of reach of current and near future experiments. This theoretical constraint can be evaded once the model includes at least two heavy neutrinos, introducing textures in the neutrino mass matrix that reproduce light neutrino masses and permit the mixing to be significantly enhanced~\cite{Casas:2001sr}. Thus, the aforementioned searches for $N_h$, which generally interpret their results in terms of one heavy neutrino with large mixing, could be considered as probing Seesaw scenarios with several heavy neutrinos, but with only one of them with a mass within the reach of the experiment.

Unfortunately, this view is not acceptable. When heavy neutrinos have enhanced mixing and large splitting between their masses, the mass matrix has strong cancellations between its elements, induces large contributions to neutrinoless double beta decay ($0\nu\beta\beta$), and leads to unacceptable quantum corrections to light neutrino masses~\cite{Ibarra:2010xw,AristizabalSierra:2011mn,Mitra:2011qr,Lopez-Pavon:2012yda,Gago:2015vma,Lopez-Pavon:2015cga,Hernandez:2018cgc,Bolton:2019pcu}. Even though the cancellations in the mass matrix can be justified by the presence of a lepton number (LN) symmetry, whose breaking generates the light neutrino masses~\cite{Branco:1988ex,Shaposhnikov:2006nn,Kersten:2007vk,Gavela:2009cd}, the constraints by $0\nu\beta\beta$ and loop corrections can only be avoided if, in addition, the heavy neutrinos appear in almost degenerate pairs at tree level, usually called pseudo-Dirac neutrinos. The reason for this is that the mass splitting is connected to new sources of lepton number violation (LNV), which at tree level do not participate in the generation of light neutrino masses. Thus, the bounds on the mass splittings suggest that searches for single $N_h$ would not be theoretically well motivated, at least from the Seesaw perspective.

An important effect of having pseudo-Dirac heavy neutrinos is that all LNV effects would be heavily suppressed, particularly for large $N_h$ masses. This brings the need of phenomenological reinterpretations of collider searches~\cite{Drewes:2019byd,Tastet:2021vwp,Abada:2022wvh}, which generally give rise to modifications of the reported bounds.

It must be noted that the bounds coming from loop corrections are theoretical. In principle, it is possible to fine-tune the light neutrino tree-level masses, such that the physical masses are correctly reproduced. Thus, these constraint are based on the desire to avoid fine-tuning between the tree and loop level contributions to physical masses. In this sense, an intriguing option was presented in~\cite{CandiadaSilva:2020hxj}, in the context of a supersymmetric extension of the Type-I Seesaw. Here, light neutrino masses were generated radiatively, with contributions from both heavy neutrinos and sneutrinos. An interesting conclusion was that large LNV parameters are still allowed in the model, as the new sneutrino loops can help to keep the corrections under control. The origin of this ``SUSY screening" effect allegedly stems from remnants of the SUSY non-renormalisation theorems~\cite{Grisaru:1979wc,Seiberg:1993vc}. 

This result has interesting implications in our discussion on searches for single heavy neutrinos, regardless of having radiative light neutrino masses or not. If sizeable LNV is permitted, it would be possible to relax the constraints on $N_h$ mass splittings\footnote{Large mass splittings would still need to be compatible with $0\nu\beta\beta$.}, allowing a straightforward interpretation of experimental results. Furthermore, the discovery of a single $N_h$ could also be interpreted as a hint in favour of supersymmetry. Thus, we consider important to further examine the findings of~\cite{CandiadaSilva:2020hxj} in our context. In addition, we consider that a more detailed explanation of the screening effect is necessary, understanding which SUSY contribution allows for cancellations, and under which circumstances this happens.

In this work we take the supersymmetric extension of the Type-I Seesaw, and explore in depth the possibility of having destructive interference between the SUSY and non-SUSY loop corrections to light neutrino masses, with the intention of allowing large heavy neutrino mixing with large mass splitting. We begin by reviewing the problem of quantum corrections in Section~\ref{sec:standardseesaw}. Then, in Section~\ref{sec:susymodel}, we present the $\nu_R$MSSM and calculate the SUSY and non-SUSY loop contributions. Section~\ref{sec:Cancellations} is the most important part of this work, where we evaluate when is it feasible to have cancellations between SUSY and non-SUSY loops. We conclude in Section~\ref{sec:Conclusions} where, given our findings, we argue that due to the experimental constraints on SUSY masses the screening is not a generic feature of supersymmetry, and actually happens in very specific scenarios.

\section{Loop Corrections in the Standard Seesaw}
\label{sec:standardseesaw}

The Type I Seesaw models generate light neutrino masses via the introduction of $N$ new heavy neutral leptons $\nu_R$. These are also called \textit{sterile} neutrinos, in contrast to the \textit{active} neutrinos within $SU(2)_L$ doublets. In the model, the full neutrino mass matrix on the active-sterile basis is:
\begin{equation}
\label{eq:massmatrixtree}
M^{\rm tree}_\nu=\begin{pmatrix}
0 & M_D \\
M_D^T & M_R 
\end{pmatrix}~.
\end{equation} 
For ``large" $M_R$ one can obtain the light neutrino masses to an excellent approximation by diagonalising the matrix:
\begin{equation}
\label{eq:m_tree_approx}
M^{\rm tree}_{\rm light}=-M_D M^{-1}_R M^T_D~.
\end{equation}

On the Standard Seesaw model, the heavy neutrinos couple to Standard Model particles via the mixing matrix $U$, which diagonalises the full mass matrix shown in Eq.~(\ref{eq:massmatrixtree}). When including $N=3$ sterile neutrinos, this matrix can be decomposed into four $3\times3$ blocks:
\begin{equation}
U=\begin{pmatrix}
U_{a\ell} & U_{ah} \\
U_{s\ell} & U_{sh}
\end{pmatrix}~.
\end{equation}
Throughout this paper, $a$ indices denote the active basis where the charged lepton Yukawas $Y_e$ are diagonal, i.e.\ $a=e,\,\mu,\,\tau$. The $s=s_1,\,s_2,\,s_3$ indices denote the sterile neutrino basis, which at this point is arbitrary. In addition, $\ell=1,2,3$ labels the three light (mostly active) neutrinos $n_\ell$, with masses $m_1,\,m_2,\,m_3$, while $h=4,5,6$ labels the three heavier (mostly sterile) neutrinos $N_h$, with masses $M_4,\,M_5,\,M_6$. 

For our numerical results, we shall take a specific choice of parameters such that, in the case of normal ordering of light neutrino masses, we can write the $U_{ah}$ mixing as~\cite{Donini:2012tt,Gago:2015vma,Cerna-Velazco:2017cmn,Jones-Perez:2019plk}:
\begin{eqnarray}
\label{eq:Ua4simple}
U_{a4}
&=& i\,(U_{\rm PMNS})_{a1}\sqrt{\frac{m_1}{M_4}}~, \\
\label{eq:Ua5simple}
U_{a5}
&=& z_{56}\,Z_a\sqrt{\frac{m_3}{M_5}}\cosh\gamma_{56}\,e^{i\,z_{56}\,\rho_{56}}~, \\
\label{eq:Ua6simple}
U_{a6}
&=& i\,Z_a\,\sqrt{\frac{m_3}{M_6}}\cosh\gamma_{56}\,e^{i\,z_{56}\,\rho_{56}}~, \\
Z_a &=& (U_{\rm PMNS})_{a3}+i\,z_{56}\,\sqrt{\frac{m_2}{m_3}}(U_{\rm PMNS})_{a2}~,
\end{eqnarray}
where $z_{56}$ is the sign of the free parameter $\gamma_{56}\gtrsim2$, and $\rho_{56}\in[0,\,\pi/2]$.  From here it is possible to reconstruct the Dirac and Majorana masses appearing in Eq.~(\ref{eq:massmatrixtree}). If we take $\hat M_h={\rm diag}(M_4,\,M_5,\,M_6)$, we can write $M_D =U_{ah}^* \hat M_h$ and $M_R=\hat M_h$.

We see that both $U_{a5}$ and $U_{a6}$ can be enhanced, in this case by a factor $\cosh\gamma_{56}$, while $U_{a4}$ remains small. Thus, by taking a very large $M_4$ we can decouple this heavy neutrino, leaving us with an effective $3+2$ Seesaw model. As mentioned in the Introduction, this possibility of enhancing the active-heavy mixing while keeping acceptable light neutrino masses can be attributed to a slightly broken lepton number symmetry~\cite{Branco:1988ex,Shaposhnikov:2006nn,Kersten:2007vk,Gavela:2009cd}. 

Loop corrections can modify both $M_D$ and $M_R$, as well as generate a non-zero element in the active-active region of $M_\nu^{\rm tree}$, which can be denoted by $\delta M_D$, $\delta M_R$ and $\delta M_L$, respectively. Nevertheless, from these the most important correction to light neutrino masses comes from $\delta M_L$, such that one can write:
\begin{equation}
\label{eq:m_full_approx}
M_{\rm light}^{\rm full}=M_{\rm light}^{\rm tree}+\delta M_L~.
\end{equation}
In this Standard Seesaw model, $\delta M_L$ is determined by loops involving the $Z$ and $H^0$ bosons. Diagrams including the $W$ boson would not contribute at one loop, as there would be no LNV term on any vertex or propagator. The well-known result for $\delta M_L$~\cite{Grimus:1989pu,Pilaftsis:1991ug,Grimus:2002nk} can be written in our notation:
\begin{eqnarray}
\label{eq:deltaML}
(\delta M_L)_{aa'}&=&\dfrac{1}{v^2_{\rm SM}}\sum_{h,s,s'}(M_D)_{as}\,U_{sh}\, (M_D)_{a's'}\,U_{s'h}\, f(M_h) \\
\label{eq:deltaMLapprox}
&\approx&\frac{m_3}{v^2_{\rm SM}}\, Z_a^*Z_{a'}^*\left[M_5\,f(M_5)-M_6\,f(M_6)\right]\cosh^2\gamma_{56}\,e^{-2i\,z_{56}\,\rho_{56}}~,
\end{eqnarray}
where the loop function $f(M_h)$ is defined:
\begin{equation}
\label{eq:loopfunc}
f(M_h)=\frac{M_h}{16\pi^2}\left[3\left(\dfrac{M_h^2}{M^2_Z} - 1\right)^{-1}\ln{\dfrac{M_h^2}{M^2_Z} }+\left(\dfrac{M_h^2}{M^2_H} - 1\right)^{-1}\ln{\dfrac{M_h^2}{M^2_H} }\right]~.
\end{equation}
In Eq.~(\ref{eq:deltaMLapprox}) we have written the correction in our benchmark scenario, Eqs.~(\ref{eq:Ua4simple})-(\ref{eq:Ua6simple}), neglecting the contribution of $N_4$.

\begin{figure}[tp]
\includegraphics[width=0.45\linewidth]{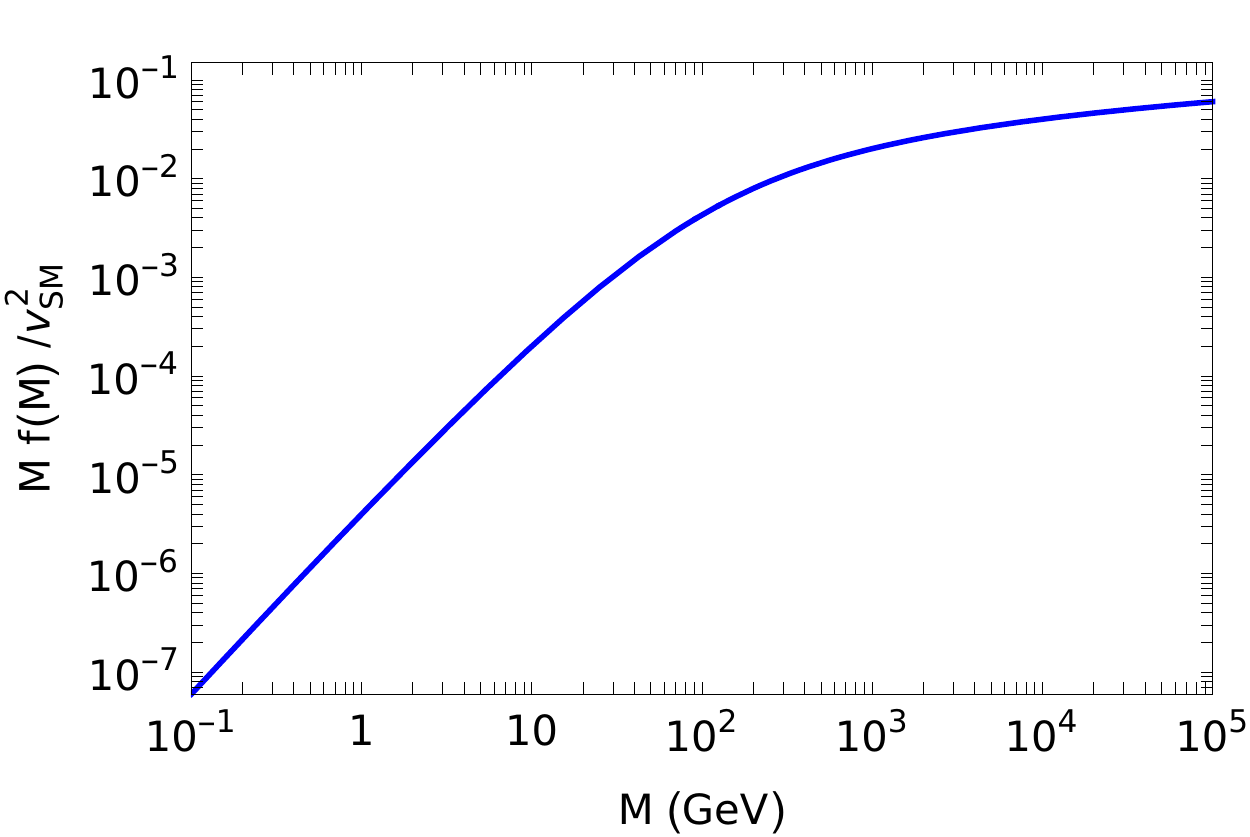} \;
\includegraphics[width=0.45\linewidth]{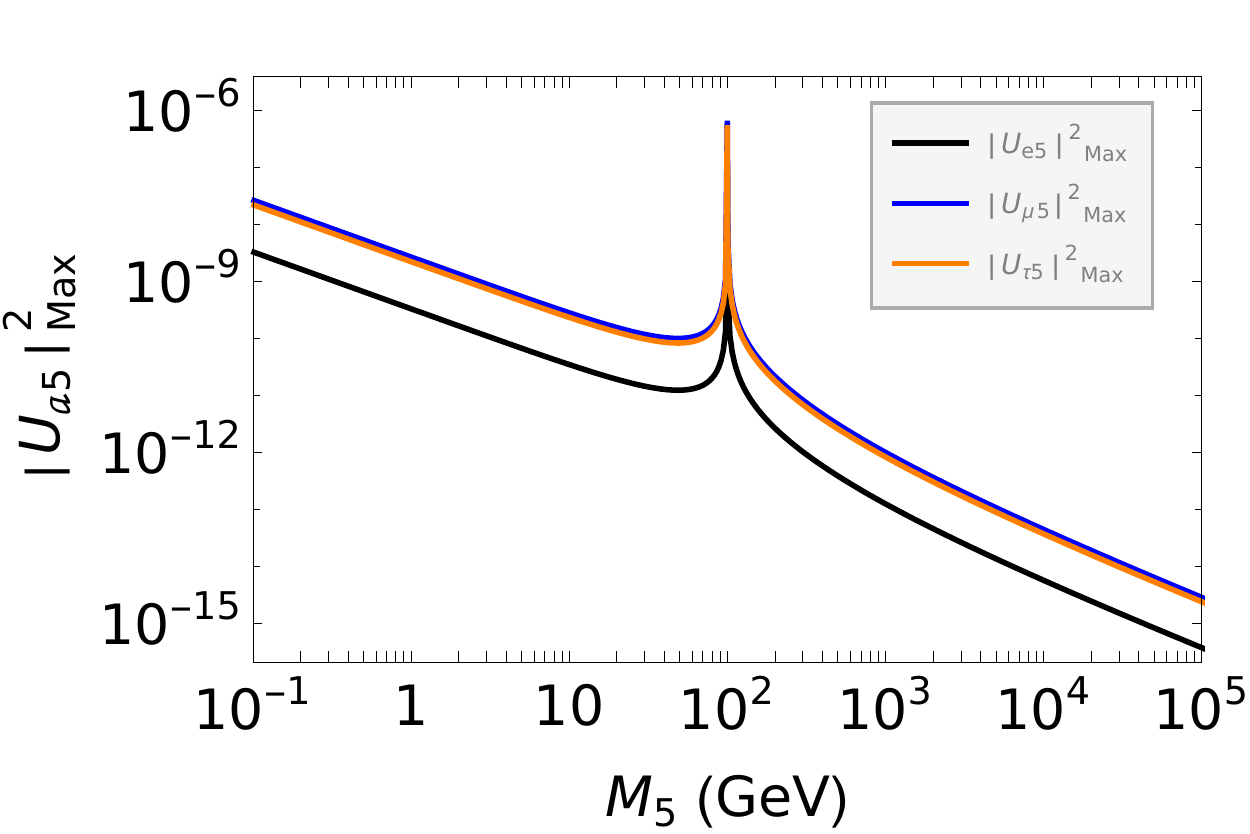}
\caption{Left: Dependence of loop function, conveniently normalised, with respect to heavy neutrino mass. Right: maximum value of $|U_{a5}|^2$ as a function of $M_5$, for $M_6=100$~GeV.}
\label{fig:loopfunction}
\end{figure}
The dependence of $f(M_h)$ as a function of the heavy neutrino mass can be seen on the left panel of Figure~\ref{fig:loopfunction}, where we have multiplied a normalisation factor $M_h/v^2_{\rm SM}$. We see the loop correction increases with mass, with the slope varying around $M_h\sim100$~GeV. This change is due to the terms multiplying the logarithms in Eq.~(\ref{eq:loopfunc}), which for large $M_h$ adds an additional suppression factor\footnote{Note that, when $M_h$ is much larger than the electroweak scale, one should actually decouple the heavy neutrinos and use effective operators.}.

From Eq.~(\ref{eq:deltaML}) it is possible to confirm that, if a heavy neutrino does not have an almost degenerate pair, then active-heavy mixing cannot exceed a certain value, or else substantial loop corrections are induced. If this bound is not respected, fine-tuning is required to accurately reproduce the observed neutrino masses~\cite{Lopez-Pavon:2015cga}. Such upper limits for $|U_{a5}|^2$ are shown as a function of $M_5$ on the right panel of Figure~\ref{fig:loopfunction}, for $M_6=100$~GeV, where we require loop corrections not to exceed $50\%$ of the tree-level value. For example, for $M_5$ equal to 1 GeV (1 TeV), we need $\gamma_{56}\lesssim2.9$ ($\lesssim2.4$), which corresponds to $|U_{\mu 5}|^2\lesssim 2 \times10^{-9}$ ($|U_{\mu 5}|^2\lesssim 8 \times 10^{-13}$). Note that the apparent stronger bounds on $|U_{e5}|^2$ are really due to the correlations existing between the mixings such that, given some value for $|U_{\mu h}|^2$ or $|U_{\tau h}|^2$, the different $Z_a$ terms make $|U_{e h}|^2$ smaller. From this result, it is clear that a single heavy neutrino with mass $\gtrsim1$~GeV cannot have its mixing enhanced by too much, so is unlikely to appear at collider searches\footnote{This statement is made evident by comparing our limits with experimental bounds shown in~\cite{Atre:2009rg,Deppisch:2015qwa,Abdullahi:2022jlv}.}.

As a final comment, note that in Eq.~(\ref{eq:deltaMLapprox}) one can see that, if $M_5\to M_6$, there exists a cancellation between the $N_5$ and $N_6$ contributions. This leads to the peak shown in the right panel of Figure~\ref{fig:loopfunction}. As commented earlier, this can again be attributed to the slightly broken lepton number symmetry, which guarantees that loop corrections are kept small~\cite{Lopez-Pavon:2012yda,Lopez-Pavon:2015cga,Hernandez:2018cgc}. In this interpretation, degenerate $N_h$ masses imply that the only non-zero sources of LNV are those essential for obtaining non-zero light neutrino masses, so no new LNV terms appear at the loop level. The maximum size of allowed non-degeneracy is critically dependent on the value of $|U_{ah}|^2$ and the average mass, as was shown in~\cite{Hernandez:2018cgc}.

\section{The $\nu_R$MSSM Model}
\label{sec:susymodel}

The simplest SUSY extension of the Standard Seesaw consists of introducing $\hat\nu^c_R$ superfields to the MSSM. Apart from the sterile neutrinos, this also implies the presence of new scalar partners, the R-sneutrinos $\tilde\nu^c_R$. The introduction of SUSY leads to modifications in the light neutrino phenomenology, for example, due to RGEs~\cite{Chankowski:1993tx,Babu:1993qv,Antusch:2005gp,Bustamante:2010bf,Singh:2018cxy}. Of course, here we are interested in the new contributions to the loop corrections to the neutrino propagator. These can be be either supersymmetric or non-supersymmetric, the former including loops with neutralinos and sneutrinos, as well as well as charginos and charged sleptons, and the latter involving the heavier Higgs bosons.

The Superpotential of the model is: 
\begin{align}
\label{eq:superpotential}
\mathcal{W} = \mathcal{W}_{\rm MSSM}
+ (Y^*_\nu)_{as}\,\hat{L}_a \cdot \hat{H}_u\, \hat{\nu}^c_{Rs}
+ \tfrac{1}{2} (M_R)_{ss'}\,\hat{\nu}^c_{Rs}\,\hat{\nu}^c_{Rs'}~.
\end{align}
where the Yukawas are connected to the Dirac mass via $M_D=\tfrac{v_{SM}}{\sqrt2} Y_\nu^*\sin\beta$, with $\tan\beta$ being the ratio of the Higgs vevs. Note that the parametrisation we are using in the neutrino sector determines $M_D$, from which the Yukawas can be extracted. In addition to the Superpotential, the following soft SUSY-breaking terms are allowed:
\begin{equation}
\mathcal{V}^{soft} =\mathcal{V}_{\rm MSSM}^{soft}
  + (m^2_{\tilde\nu})_{ss'}\tilde{\nu}^{c\,*}_{Rs}\,\tilde{\nu}^c_{Rs'}
  + \left(\tfrac{1}{2}(B_{\nu})_{ss'}\tilde{\nu}^c_{Rs}\,\tilde{\nu}^c_{Rs'} + (T^*_\nu)_{as}\,\tilde{L}_a \cdot H_u\, \tilde{\nu}^c_{Rs} + \text{H.c.} \right)~.
\end{equation}
In addition to the typical soft mass $m_{\tilde\nu}^2$ and trilinear couplings $T_\nu$, we have a LNV soft mass $B_\nu$. This new term will give further contributions to neutrino masses at the loop level. In fact, $B_\nu$ played a major role in~\cite{Hirsch:2009ra}, in the context of the supersymmetric inverse seesaw with only one pair of sterile neutrinos. Here, they explored the possibility of generating one light neutrino mass via the Standard Seesaw, and the other through SUSY corrections, with the requirement of having $T_\nu$ not aligned with $Y_\nu$.

In the following sections we shall describe both SUSY and non-SUSY loop corrections to the light neutrino masses. Here, and in the following Sections, we will focus on heavy neutrino masses discoverable at colliders, namely $M_h=40,\,200$~GeV.

\subsection{Non-SUSY Loop Corrections}
\label{sec:non-susy}

\begin{figure}[tb]
\includegraphics[width=0.4\linewidth]{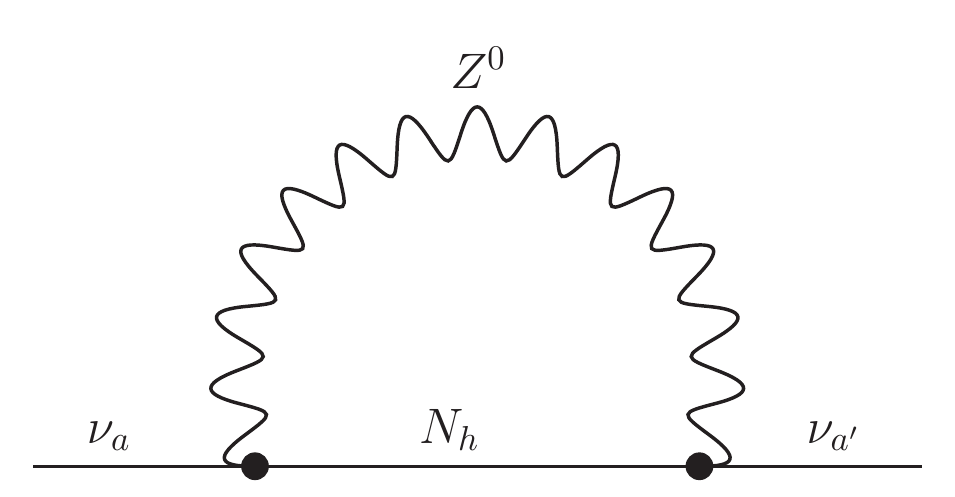}\hspace{10mm}
\includegraphics[width=0.4\linewidth]{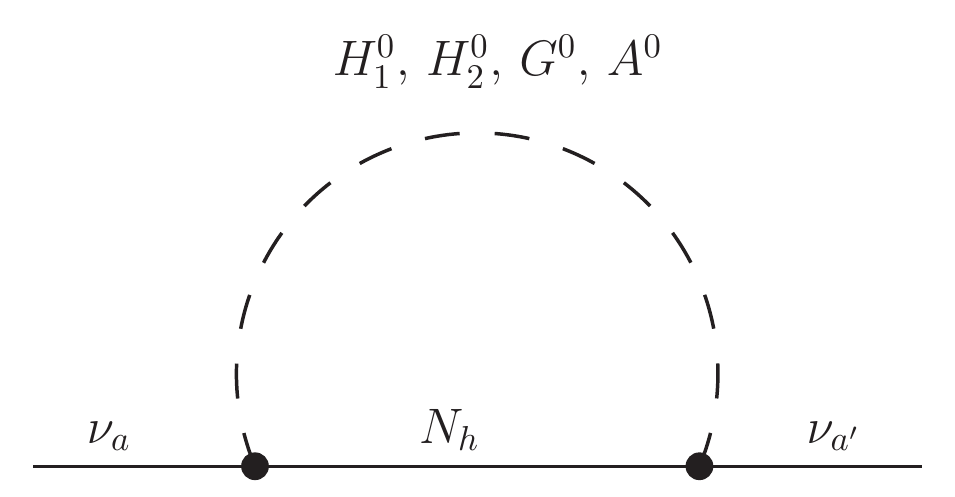}
\caption{Non-supersymmetric one-loop diagrams of the MSSM that contribute to the neutrino mass matrix.}
\label{fig:2hdm}
\end{figure}
The full one-loop correction to the neutrino propagator in models with two Higgs doublets has been extensively studied in the past, see for example~\cite{Grimus:1989pu,Grimus:1999wm,Grimus:2002nk,Ibarra:2011gn,Jurciukonis:2015rha,Grimus:2018rte,Jurciukonis:2019jkr}. The relevant diagrams involve the $W$ and $Z$ bosons, the neutral and charged Higgs bosons, and the corresponding Goldstone bosons. Nevertheless, as in the Standard Seesaw, loops involving charged particles do not contribute to the Majorana mass, leaving only the diagrams shown in Figure~\ref{fig:2hdm}. Although the $\delta M_L$ correction involving the $Z$ is the same as in the Standard Seesaw, there is a new combined contribution from the neutral scalars. Thus, in terms of $Y_\nu$, we can write the full correction:
\begin{eqnarray}
(\delta M_L)^{\rm 2HDM}_{aa'}&=&\dfrac{1}{2}\sum_{h,s,s'}(Y^*_\nu)_{as}\,U_{sh}\, (Y^*_\nu)_{a's'}\,U_{s'h}\, g(M_h,\,M_A,\,\tan\beta) \\
\label{eq:deltaML2HDM}
&\approx& K_{aa'} \left[M_5\,g(M_5,\,M_A,\,\tan\beta)-M_6\,g(M_6,\,M_A,\,\tan\beta)\right]~,
\end{eqnarray}
where the second line again corresponds to our benchmark scenario. We have defined $K_{aa'}=(m_3/v_u^2)\,
Z_a^*Z_{a'}^*\cosh^2\gamma_{56}\,e^{-2iz_{56}\rho_{56}}$, and a new loop function:
\begin{multline}
\label{eq:loopfunc1}
g(M_h,\,M_A,\,\tan\beta)=\frac{M_h}{16\pi^2}\left[3\sin^2\beta\left(\dfrac{M_h^2}{M^2_Z} - 1\right)^{-1}\ln{\dfrac{M_h^2}{M^2_Z} }
+\cos^2\alpha\left(\dfrac{M_h^2}{M^2_{H_1}} - 1\right)^{-1}\ln{\dfrac{M_h^2}{M^2_{H_1}} }\right. \\
\left.
+\sin^2\alpha\left(\dfrac{M_h^2}{M^2_{H_2}} - 1\right)^{-1}\ln{\dfrac{M_h^2}{M^2_{H_2}}}
-\cos^2\beta\left(\dfrac{M_h^2}{M^2_{A}} - 1\right)^{-1}\ln{\dfrac{M_h^2}{M^2_{A}}}
\right]~.
\end{multline}
Here, $M_A$, $M_{H_1}$ and $M_{H_2}$ are the masses of the pseudoscalar and scalar Higgses, and $\alpha$ is the scalar mixing angle. It is important to remember that, at tree level, all of the latter are a function of $M_A$ and $\tan\beta$. In particular, in the decoupling regime, we find a very precise cancellation between the $H_2$ and $A$ contributions. Notice we do not proceed as in~\cite{CandiadaSilva:2020hxj}, who modify the effective quartic coupling in the scalar mass matrix such that the observed lightest Higgs mass is obtained. The reason is that the aforementioned cancellation is spoilt, suggesting it might be necessary to include similar corrections in the pseudoscalar mass matrix at the same time, which is outside the scope of this work.

\begin{figure}[tp]
\includegraphics[width=0.45\linewidth]{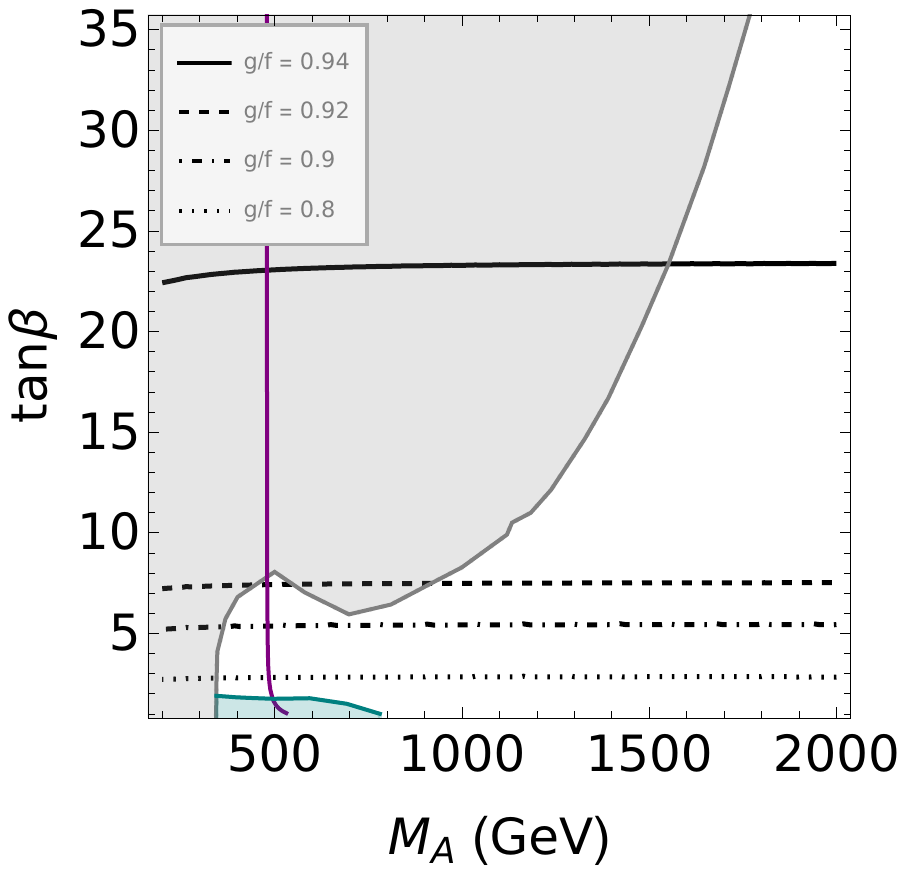}\hfill
\includegraphics[width=0.45\linewidth]{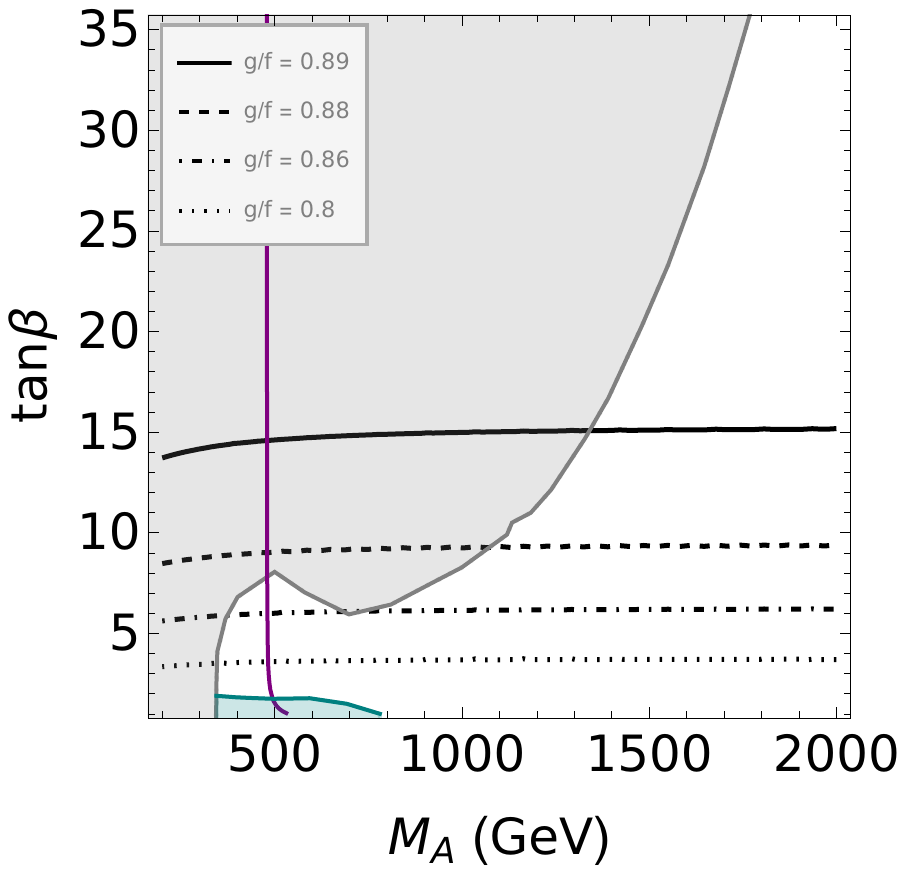}
\caption{Ratio between loop functions, $g(M_h,\,M_A,\,\tan\beta)/f(M_h)$. Gray (dark green) region is excluded by $H/A\to\tau^+\tau^-$~\cite{ATLAS:2020zms} ($H^\pm\to tb$~\cite{ATLAS:2021upq}) searches. The region to the left of the purple curve is excluded, as here the light Higgs boson couplings do not match with measurements~\cite{ATLAS:2019nkf}. We show results for $M_h=40$ ($200$)~GeV on the left (right).}
\label{Fig:gOverf}
\end{figure}
From Eq.~(\ref{eq:deltaML2HDM}), we can expect that corrections in general Type-II two Higgs doublet models will have a very similar phenomenology to that of the Standard Seesaw, in particular in what concerns the enhancement to the neutrino mixing and the possibility of cancellations between different heavy neutrino contributions. In our case, the constraints imposed by the SUSY framework appear in the structure of the $g$ function, as shown in Eq.~(\ref{eq:loopfunc1}), where most of the appearing parameters are related to each other. This leads to $g$ not having a strong dependence on $M_A$ nor $\tan\beta$, with numerical values very similar to the $f$ of the Standard Seesaw, see Eq.~(\ref{eq:loopfunc}). To illustrate this, we shown in Figure~\ref{Fig:gOverf} the ratio between $g$ and $f$, presented as a function of $M_A$ and $\tan\beta$, for two values of $M_h$. We find that, for the evaluated values of heavy neutrino mass, $g$ is always slightly smaller than $f$, but hardly decreases under $80\%$.

Thus, for a given $M_D$ and $\hat M_h$, the non-SUSY corrections are expected to be of the same order of magnitude as in the Standard Seesaw. These will depend on the heavy neutrino masses in a way similar to what is shown on the left panel of Figure~\ref{fig:loopfunction}. Correspondingly, the larger the $\Delta M_{65}=M_6-M_5$ mass splitting, the larger the contribution, with its sign being the opposite of that of $\Delta M_{65}$.

\subsection{SUSY Loop Corrections}
\label{sec:susyloop}

As mentioned earlier, SUSY corrections to the light neutrino propagator involve both sneutrino - neutralino and charged slepton - chargino loops. Moreover, since only the $\hat\nu_R$ sector involve LNV terms, only the former are relevant for $\delta M_L$~\cite{Dedes:2007ef,Hollik:2014hya,CandiadaSilva:2020hxj}.

Since we now have two sources of LNV, namely $M_R$ and $B_{\nu}$, for transparency we will carry out our analysis using the mass-insertion technique~\cite{Gabbiani:1996hi,Misiak:1997ei,Hisano:1998fj,Raz:2002zx, Dedes:2015twa}. This has the additional advantage of being able to carry out our calculations directly on the active-sterile basis. Such an approach was also followed in~\cite{CandiadaSilva:2020hxj}, although here this will be done only for the sneutrino line in the SUSY contribution. For this, we need to write the terms of the sneutrino scalar potential contributing to the sneutrino mass matrix. These can be split into LNC and LNV terms, $\mathcal L^{\rm mass}_{\tilde\nu}=\mathcal L^{\rm LNC}_{\tilde\nu}+\mathcal L^{\rm LNV}_{\tilde\nu}$, where:
\begin{eqnarray}
\label{eq:SUSY_LNC_Lag}
-\mathcal L^{\rm LNC}_{\tilde\nu}&=& 
\tilde\nu_{La}^*\underbrace{\left(m^2_{\tilde L} +\frac{v_u^2}{2}Y_\nu\,Y_\nu^\dagger +\frac{1}{2}m_Z^2\cos2\beta\right)_{aa'}}_{m^2_{\tilde\nu_L}}\tilde\nu_{La'} \nonumber \\
&&+ \tilde\nu^{c}_{Rs}\underbrace{\left(m^{2\,T}_{\tilde\nu}
+\frac{v_u^2}{2}Y^\dagger_\nu\,Y_\nu +M_R\,M^*_R\right)_{ss'}}_{m^2_{\tilde\nu_R}}\tilde\nu^{c\,*}_{Rs'} \nonumber \\
&&+\tilde\nu^c_{Rs}\left(\frac{v_u}{\sqrt2}T^\dagger_\nu -\frac{v_d}{\sqrt2}\mu^* Y^\dagger_\nu\right)_{sa}\tilde\nu_{La}
+\tilde\nu^*_{La}\left(\frac{v_u}{\sqrt2}T_\nu -\frac{v_d}{\sqrt2}\mu Y_\nu\right)_{as}\tilde\nu^{c\,*}_{Rs}\\
\label{eq:SUSY_LNV_Lag}
-\mathcal L^{\rm LNV}_{\tilde\nu}&=&
\tilde\nu^{c}_{Rs}\left(\frac{1}{2}\,B_\nu\right)_{ss'}\tilde\nu^{c}_{Rs'}
+\tilde\nu^*_{La}\left(\frac{v_u}{\sqrt2}Y_\nu\,M_R\right)_{as}\tilde\nu^{c}_{Rs}
+\tilde\nu_{La}\left(\frac{v_u}{\sqrt2}Y^*_\nu\,M^*_R\right)_{as}\tilde\nu^{c\,*}_{Rs}
\end{eqnarray}
Thus, we have LNV mass insertions from Eq.~(\ref{eq:SUSY_LNV_Lag}), as well as LNC insertions from the last line of Eq.~(\ref{eq:SUSY_LNC_Lag}). From Eq.~(\ref{eq:SUSY_LNV_Lag}), we find two types of LNV terms. From these, the $Y_\nu M_R$ terms are ``irreducible" in the sense that they cannot be set to zero without spoiling the seesaw mechanism at tree level. In contrast, a vanishing $B_\nu$ does not affect the neutrino masses at leading order, and thus are considered ``reducible".

In the following, for simplicity, we assume $m_{\tilde L}^2$ and $m^2_{\tilde\nu}$ to be diagonal. With this, we can also take the $m^2_{\tilde\nu_L}$ and $m^2_{\tilde\nu_R}$ matrices as diagonal, to an excellent approximation. In addition, when presenting numerical results, we will take $T_\nu=a_\nu\,Y_\nu$ and $B_\nu=b_\nu\,M_R$. Note that these assumptions, which are not guaranteed by SUSY, will be crucial to preserve the flavour structure of the tree-level mass matrix.

In what follows, we list all possible contributions to $\delta M_L$ up to order $\ord{Y_\nu^2}$, which was the assumption taken when writing Eq.~(\ref{eq:m_full_approx}). For each type of loop diagram, we present both the complete expression and an approximate one relevant for our benchmark scenario, applying our assumptions for $T_\nu$ and $B_\nu$, taking degenerate\footnote{Exactly degenerate sleptons can induce artificially large mixing. This can be avoided by adding slepton mass splittings at the per-mille level, without spoiling our numerical results.} $m_{\tilde L}$ and neglecting the contribution from $\tilde\nu_{R4}$.

\subsubsection{Irreducible Contributions ($B_\nu=0$)}

\begin{figure}[tp]
\includegraphics[width=0.39\textwidth]{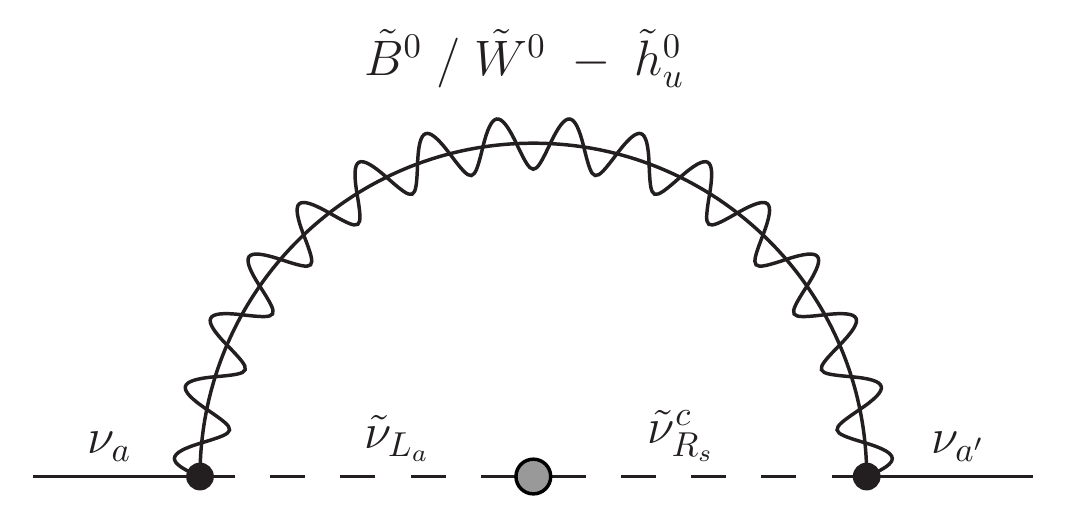}
\includegraphics[width=0.39\textwidth]{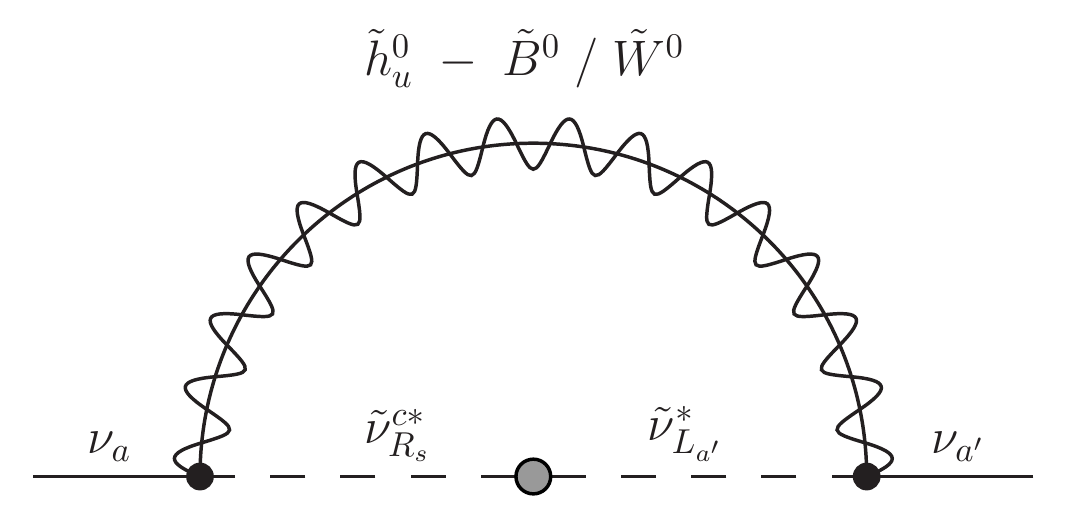}\\[1em]
\includegraphics[width=0.39\textwidth]{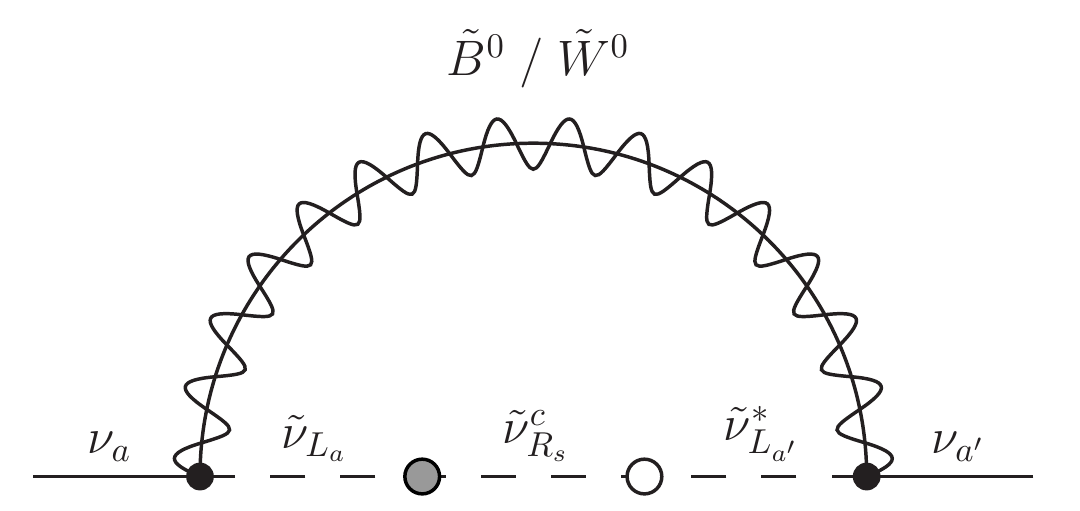}
\includegraphics[width=0.39\textwidth]{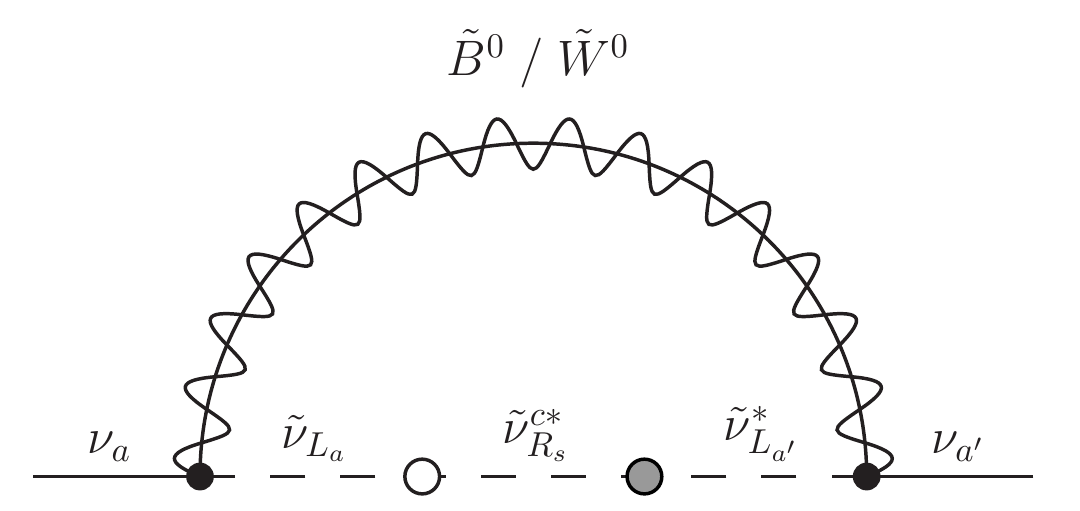} 
\caption{Irreducible mass insertions.  In all cases, gray (white) blobs indicate LNV (LNC) insertions. Top: gaugino - higgsino case. Bottom: pure gaugino case.}
\label{Fig:MassInsertions1}
\end{figure}
Since we are taking terms of order $\ord{Y_\nu^2}$, it is crucial to note that the LNV insertions we are currently considering are of the type $(Y_\nu\,M_R)/M_{\rm SUSY}$, meaning that we will have at most two of these in $\delta M_L$. The same reasoning can be followed for the LNC insertions in the last line of Eq.~(\ref{eq:SUSY_LNC_Lag}). From these considerations, we can expect these SUSY corrections to be negligible if the Yukawas are not enhanced.

Let us consider the pure higgsino contribution. Here, we have a $Y_\nu$ suppression at each vertex, so adding any insertion make these of order larger than $\ord{Y_\nu^2}$, and can be neglected.

\begin{figure}[tp]
\includegraphics[width=0.4\linewidth]{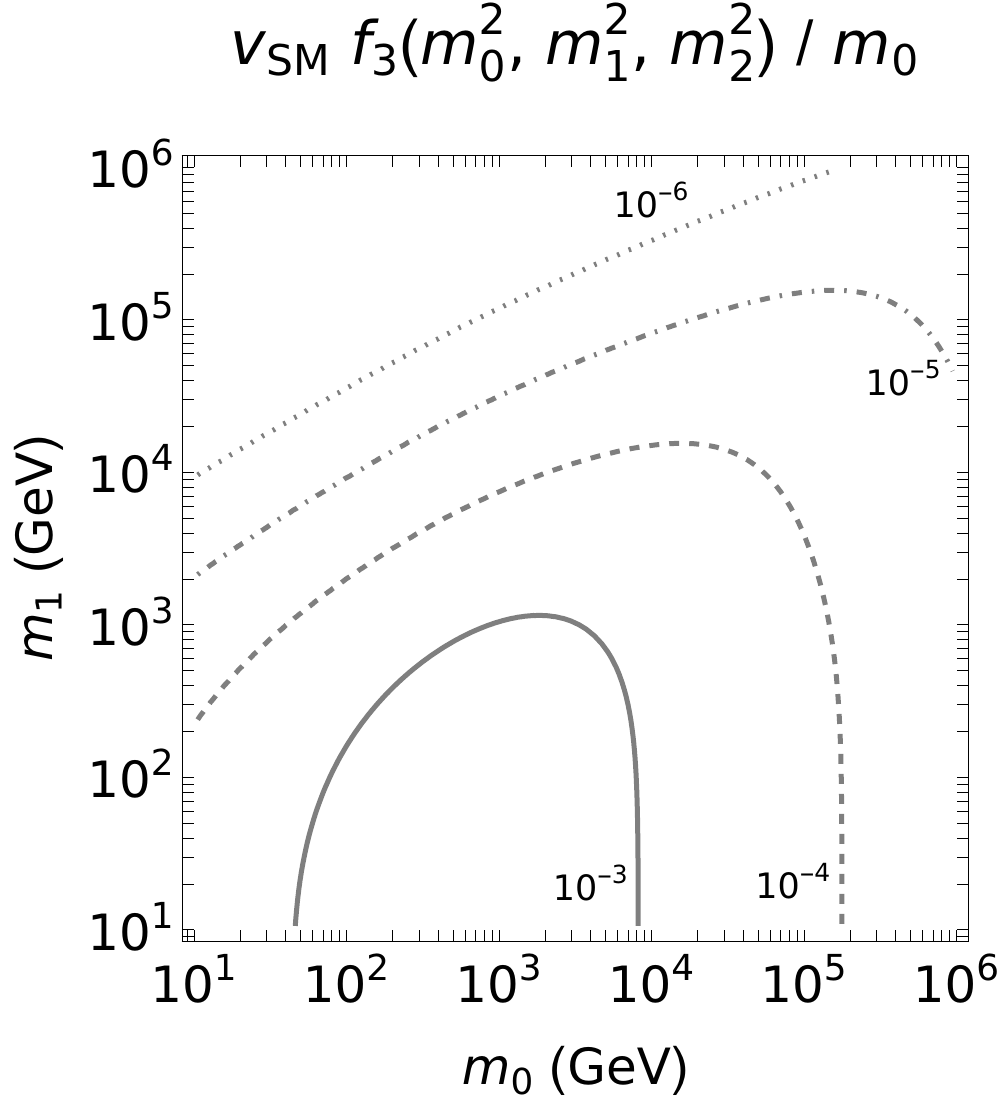}\qquad\qquad
\includegraphics[width=0.4\linewidth]{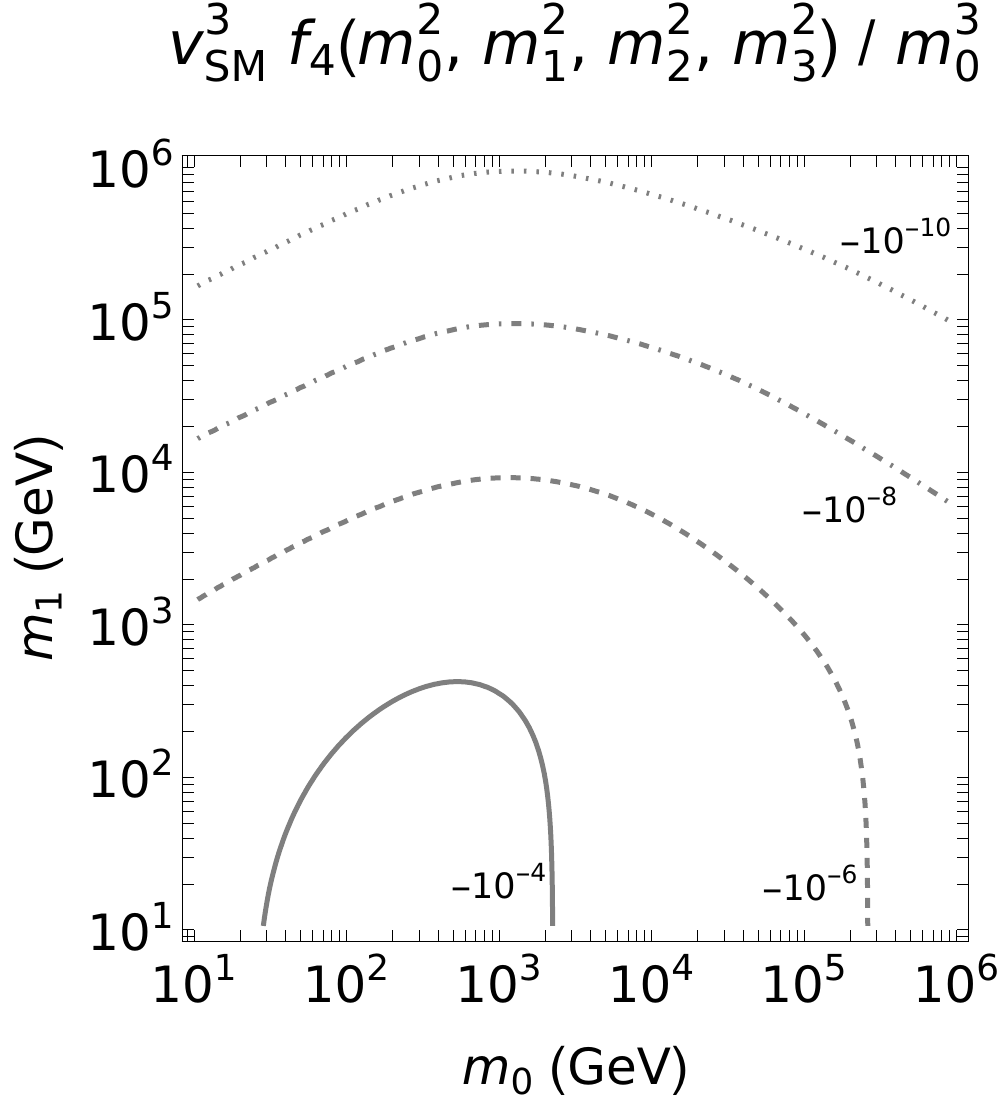}
\caption{Mass insertion functions $f_3$ (left) and $f_4$ (right), each with a convenient normalisation. We set $m_2 = m_3 = 600\,$GeV.}
\label{Fig:MIfunc4}
\end{figure}

Next come the gaugino-higgsino contributions, shown on the top row of Figure~\ref{Fig:MassInsertions1}, with only one vertex with a $Y_\nu$ suppression. We can allow only one mass-insertion:
\begin{eqnarray}
\label{eq:gaugino_higgsino1}
(\delta M^{\rm irr}_L)^{gh}_{aa'}
&=& \frac{v_u}{2}\sum_{b,s,r}(-1)^{b}g_b\frac{1}{m_{\tilde{\chi}^0_r}} O_{rb}O_{r4}\left[(Y^*_\nu M^*_R)_{as}(Y^*_{\nu})_{a's}
f_3(m^2_{\tilde{\chi}^0_r},m^2_{ \tilde{\nu}_{La}},m^2_{\tilde{\nu}_{Rs}})\right. \nonumber \\
&&\hspace{5cm}\left.+(Y^*_{\nu})_{as}(Y^*_\nu M^*_R)_{a's}
f_3(m^2_{\tilde{\chi}^0_r},m^2_{\tilde{\nu}_{Rs}},m^2_{ \tilde{\nu}_{La'}})\right] \\
\label{eq:gaugino_higgsino2}
&\approx& 2v_u\, K_{aa'} \sum_{b,r} (-1)^{b}g_b\frac{1}{m_{\tilde{\chi}^0_r}} O_{rb}O_{r4} \nonumber \\
&&\hspace{2.6cm}\times\left[M^2_5\,f_3(m_{\tilde{\chi}^0_r}^2,m^2_{\tilde\nu_{R5}},m^2_{\tilde\nu_L}) -M^2_6\, f_3(m_{\tilde{\chi}^0_r}^2,m^2_{\tilde\nu_{R6}}, m^2_{\tilde\nu_L})\right]
\end{eqnarray}
where $O_{rb}$ are the neutralino mixing matrices, $r = 1,\ldots,4$ denotes the neutralino mass eigenstates, and $b$ can be 1 (bino) or 2 (wino). The function $f_3$ is defined:
\begin{equation}
\label{eq:f3}
f_3(m^2_0,m^2_1,m^2_2)=\frac{1}{16\pi^2}\frac{m^2_0}{m^2_1-m^2_2}\left[\left(1-\frac{m^2_0}{m^2_1}\right)^{-1} \ln\frac{m^2_1}{m^2_0}-\left(1-\frac{m^2_0}{m^2_2} \right)^{-1} \ln\frac{m^2_2}{m^2_0}\right]
\end{equation}
and is shown on the left panel of Figure~\ref{Fig:MIfunc4}, being symmetric with respect to $m_1\leftrightarrow m_2$ exchange. It is clear that $f_3$ is largest when $m_1$ and $m_2$ are smallest. Furthermore, for fixed $m_1,\,m_2$, this function is maximised when $m_0\sim{\rm Max}(m_1,\,m_2)$.

Finally, for the pure gaugino case we have no suppressed vertices, but need two $LR$ transitions on the sneutrino line. As shown on the bottom row of Figure~\ref{Fig:MassInsertions1}, in order to contribute to $\delta M_L$, these must combine one LNV and one LNC mass insertion:
\begin{eqnarray}
(\delta M^{\rm irr}_L)^{gg}_{aa'} 
 &=&\frac{v^2_u}{4} \sum_{b,b',s,r}(-1)^{b+b'}g_b \,g_{b'} \frac{1}{ m^3_{\tilde{\chi}^0_r}}O_{rb}O_{rb'}  f_4(m^2_{\tilde{\chi}^0_r},m^2_{ \tilde{\nu}_{La}},m^2_{\tilde{\nu}_{R s}},m^2_{ \tilde{\nu}_{La'}}) \nonumber \\
 &&\hspace{0.5cm}\times\left[
 (T^*_\nu-\mu^*Y^*_\nu\cot{\beta})_{a s}
 (Y^*_\nu M^*_R)_{a's} +(Y^*_\nu M^*_R)_{as}
 (T^*_\nu-\mu^*Y^*_\nu\cot{\beta})_{a's}
 \right]  \\
 \label{eq:gaugino_gaugino2}
 &\approx& v^2_u\, K_{aa'} (a_\nu-\mu \cot \beta ) \sum_{b,b',r} (-1)^{b+b'}g_b\, g_{b'} \frac{1}{m^3_{\tilde{\chi}^0_r}} O_{rb}O_{rb'} \nonumber \\
&&\hspace{1.25cm}\times\left[ M^2_5\,f_4(m^2_{\tilde{\chi}^0_r},m^2_{\tilde\nu_{R5}},m^2_{\tilde\nu_L},m^2_{\tilde\nu_L}) -M^2_6\, f_4(m^2_{\tilde{\chi}^0_r},m^2_{\tilde\nu_{R6}},m^2_{\tilde\nu_L},m^2_{\tilde\nu_L})\right]
\end{eqnarray}
The function $f_4$ follows the general expression:
\begin{equation}
\label{eq:fn}
f_n(m^2_0,m^2_1,m^2_2,...,m^2_{n-1})=\frac{m^2_0}{m^2_1-m^2_2}\left[f_{n-1}(m^2_0,m^2_1,m^2_3,...,m^2_{n-1}) -f_{n-1}(m^2_0,m^2_2,m^2_3,...,m^2_{n-1})\right]
\end{equation}
where $n-1$ is the number of mass insertions in the diagram. It is shown on the right panel of Figure~\ref{Fig:MIfunc4}, and again is symmetric under exchange of $m_i\leftrightarrow m_j$, $i,j\neq0$. As with $f_3$, the function $f_4$ has larger values for smaller $m_i$, and $m_0$ around the largest of the $m_i$.

\subsubsection{Reducible Contributions ($B_\nu\neq0$)}

Once $B_\nu$ is different from zero, a new set of loop corrections can enter the game. For this it is important to assume $B_\nu$ not to be as suppressed as $Y_\nu$, allowing diagrams with a larger number of insertions.

\begin{figure}[tp]
\includegraphics[width=0.39\textwidth]{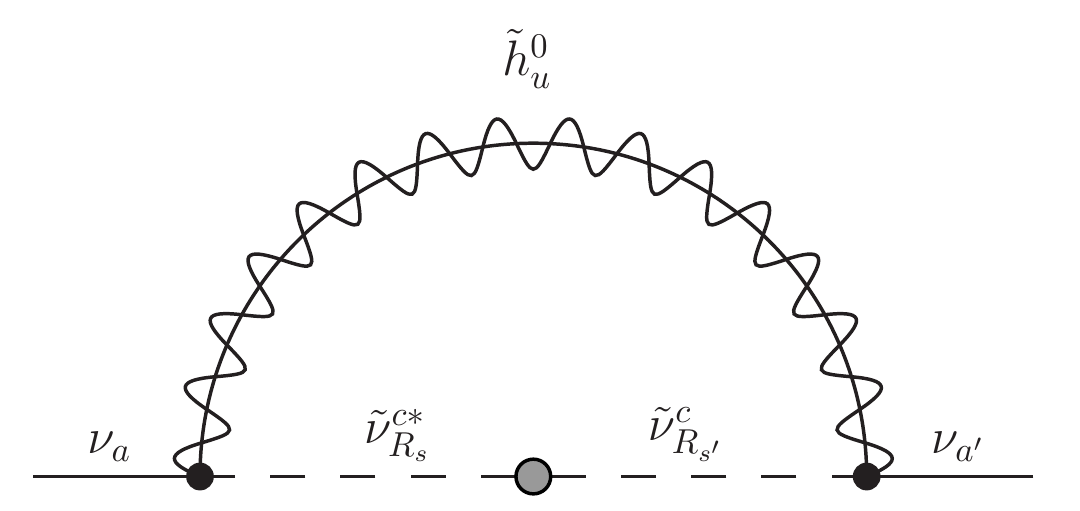}\\[1em]
\includegraphics[width=0.39\textwidth]{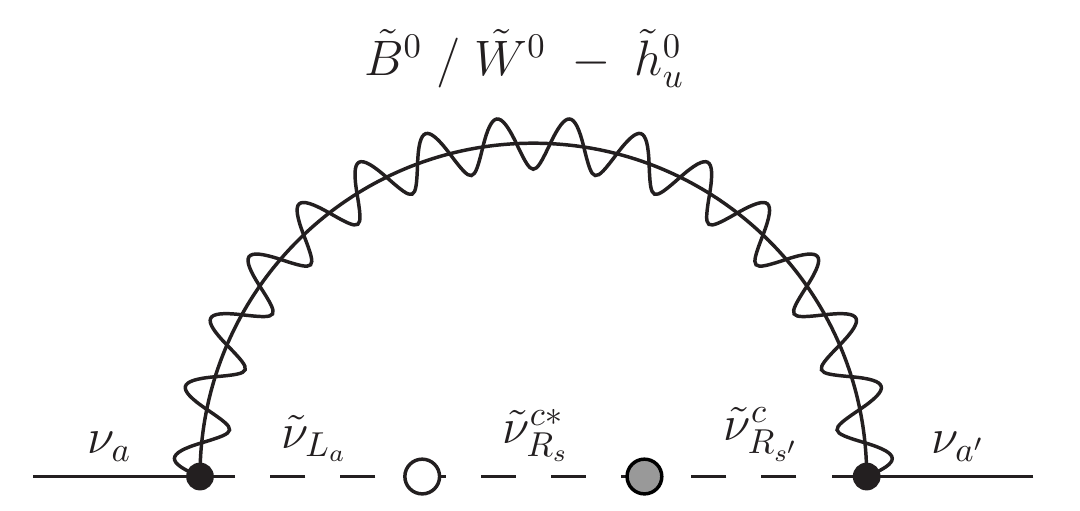}
\includegraphics[width=0.39\textwidth]{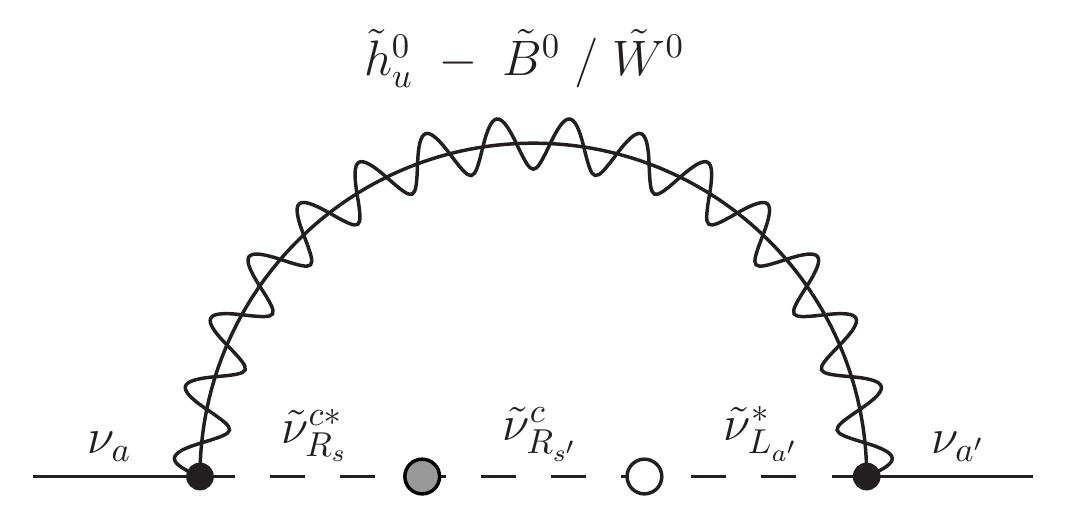} \\[1em]
\includegraphics[width=0.39\textwidth]{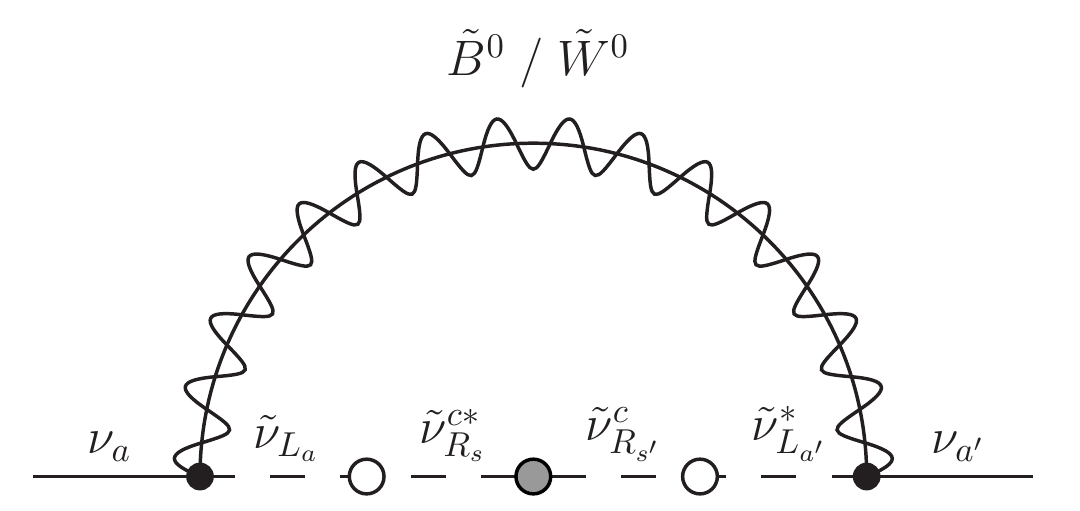}
\includegraphics[width=0.39\textwidth]{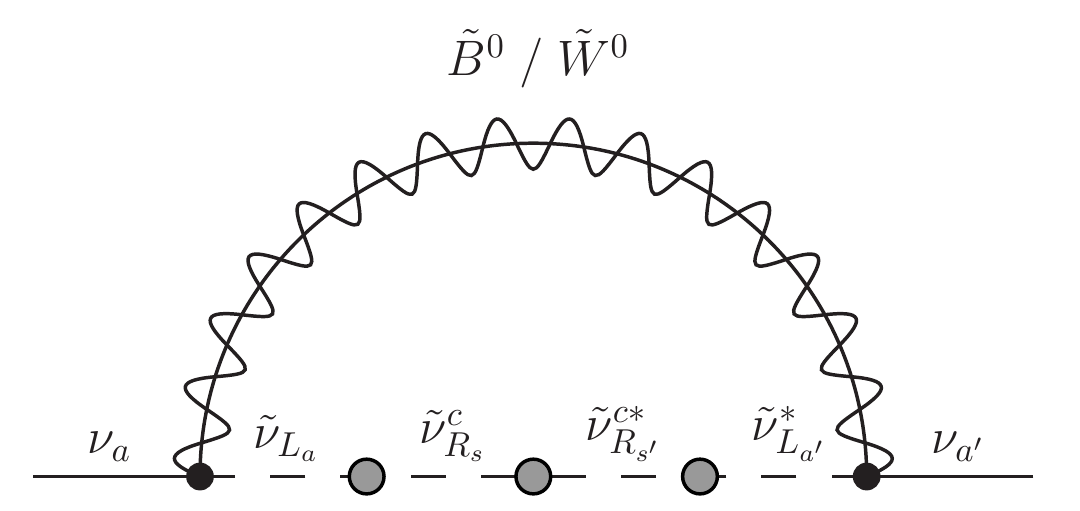}
\caption{Reducible mass insertions.  In all cases, gray (white) blobs indicate LNV (LNC) insertions. Top: pure higgisno case. Center: gaugino - higgsino case. Bottom: pure gaugino case.}
\label{Fig:MassInsertions2}
\end{figure}
Let us start again with the pure higgsino case, which had negligible irreducible contributions. This time, the self-energy, shown in the top row of Figure~\ref{Fig:MassInsertions2}, is given by:
\begin{eqnarray}
(\delta M^{\rm red}_L)^{hh}_{aa'}
&=&\sum_{r,s,s'}\dfrac{1}{m_{\tilde{\chi}^0_r}}O^2_{r4}(Y^*_{\nu})_{as}(B^*_\nu)_{ss'}(Y^*_{\nu})_{a's'}  f_3(m^2_{\tilde{\chi}^0_r},m^2_{\tilde{\nu}_{R s}},m^2_{\tilde{\nu}_{Rs'}}) \\
\label{eq:redhh}
&\approx& 2 K_{aa'}\,b_\nu 
\sum_r \frac{1}{m_{\tilde{\chi}^0_r}} O^2_{r4} 
\big[M^2_5~f_3(m_{\tilde{\chi}^0_r}^2,m^2_{\tilde\nu_{R5}},m^2_{\tilde\nu_{R5}}) 
-M^2_6f_3(m_{\tilde{\chi}^0_r}^2,m^2_{\tilde\nu_{R6}},m^2_{\tilde\nu_{R6}})\big]
\end{eqnarray}
The loop function $f_3$ is shown in Eq.~(\ref{eq:f3}), and illustrated on the left panel of Figure~\ref{Fig:MIfunc4}.

For the the gaugino-higgsino correction, we now can have two insertions, one of them being a $LR$ transition that is LNC, and then the $RR$ insertion from $B_\nu$. This can be seen on the centre row of Figure~\ref{Fig:MassInsertions2}, leading to:
\begingroup
\allowdisplaybreaks
\begin{eqnarray}
(\delta M^{\rm red}_L)^{gh}_{aa'}
 &=&  \frac{v_u}{4}\sum_{b,s,s',r}(-1)^{b}g_b
 \frac{1}{m^3_{\tilde{\chi}^0_r}}O_{rb}O_{r4} \nonumber \\
 &&\qquad\times\left[
(T^*_\nu- \mu^{*} \cot\beta ~Y^*_{\nu})_{as'} (B^{*}_{\nu})_{ss'}(Y^*_{\nu})_{a's}
f_4(m^2_{\tilde{\chi}^0_r},m^2_{ \tilde{\nu}_{La}},m^2_{\tilde{\nu}_{Rs'}},m^2_{\tilde{\nu}_{Rs}})\right. \nonumber \\
&&\qquad\left.+(Y^*_{\nu})_{as}(B^{*}_{\nu})_{s's}
(T^*_\nu- \mu^{*} \cot\beta ~Y^*_{\nu})_{a's'} 
f_4(m^2_{\tilde{\chi}^0_r},m^2_{\tilde{\nu}_{Rs}},m^2_{\tilde{\nu}_{Rs'}},m^2_{ \tilde{\nu}_{La'}})\right] \\
\label{eq:redgh}
&\approx& v_u\,K_{aa'}\,(a_\nu-\mu \cot \beta )\,b_\nu 
\sum_{b, r} (-1)^{b}g_b\frac{1}{m^3_{\tilde{\chi}^0_r}}
O_{rb}O_{r4} \nonumber \\
&&\hspace{1cm}\times
\left[M^2_5\,f_4(m^2_{\tilde{\chi}^0_r},m^2_{\tilde\nu_L},m^2_{\tilde\nu_{R5}},m^2_{\tilde\nu_{R5}}) -M^2_6\,f_4(m^2_{\tilde{\chi}^0_r},m^2_{\tilde\nu_L},m^2_{\tilde\nu_{R6}},m^2_{\tilde\nu_{R6}})\right]
\end{eqnarray}
\endgroup
Here, the loop function $f_4$ is deduced from Eq.~(\ref{eq:fn}), and shown on the right panel of Figure~\ref{Fig:MIfunc4}.

Finally, following the bottom row of Figure~\ref{Fig:MassInsertions2}, the pure gaugino loops have three mass insertions, with two $LR$ transitions in addition to the $B_\nu$ insertion:
\begin{eqnarray}
(\delta M^{\rm red}_L)^{gg}_{aa'}
 &=& \dfrac{v^2_u}{8}\sum_{b,b',s,s',r}(-1)^{b+b'} g_b\, g_{b'}
 \frac{1}{m^5_{\tilde{\chi}^0_r}} O_{rb} O_{rb'}
 \,f_5(m^2_{\tilde{\chi}^0_r},m^2_{ \tilde{\nu}_{L a}},m^2_{\tilde{\nu}_{R s}},m^2_{\tilde{\nu}_{R s'}},m^2_{ \tilde{\nu}_{La'}}) \nonumber \\
 &&\hspace{1cm}\times\left[
 (Y^*_{\nu} M^*_R)_{as}(B_{\nu})_{ss'}(Y^*_{\nu} M^*_R)_{a's'}\right. \nonumber \\
 &&\hspace{3cm}\left.+
 (T^*_{\nu}-\mu^{*}\cot\beta~ Y^*_{\nu})_{a s}
  (B^*_{\nu})_{ss'}
 (T^*_{\nu}-\mu^{*}\cot\beta~Y^*_{\nu})_{a's' }  \right] \\
\label{eq:redgg}
 &\approx& \frac{v^2_u}{4}K_{aa'}\,b_\nu\sum_{b,b',r} (-1)^{b+b'}g_b\, g_{b'}
 \frac{1}{m^5_{\tilde{\chi}^0_r}}
O_{rb}O_{rb'} \nonumber \\
&&\hspace{1cm}\times\left[M^2_5\left(M^2_5+(a_\nu-\mu \cot \beta )^2\right)f_5(m^2_{\tilde{\chi}^0_r},m^2_{\tilde\nu_{R5}},m^2_{\tilde\nu_{R5}},m^2_{\tilde\nu_L},m^2_{\tilde\nu_L})\right. \nonumber \\
&&\hspace{1.75cm}\left.-M^2_6\left(M^2_6+(a_\nu-\mu \cot \beta )^2\right) f_5(m^2_{\tilde{\chi}^0_r},m^2_{\tilde\nu_{R6}},m^2_{\tilde\nu_{R6}},m^2_{\tilde\nu_L},m^2_{\tilde\nu_L}) \right]
\end{eqnarray}
As before, the function $f_5$ is calculated following Eq.~(\ref{eq:fn}), following the general properties outlined earlier for $f_3$ and $f_4$.

\section{Searching for Cancellations}
\label{sec:Cancellations}

The main purpose of this work is to find under what conditions can one have large LNV terms while keeping loop corrections under control, avoiding the need for fine-tuning in the neutrino sector. Since we know that the non-SUSY contributions are very similar to those in the Standard Seesaw, it is necessary to have a cancellation featuring the sneutrino loops. In fact, presenting this SUSY screening was the main motivation of~\cite{CandiadaSilva:2020hxj}, where it was suggested that the non-renormalisation theorems could enforce such a result. So, in the following we concentrate on characterising the region of the parameter space guaranteeing destructive interference, for both reducible and irreducible corrections.

\subsection{Irreducible Contributions}

In order to understand how these corrections affect neutrino masses, let us first assume that $\tilde\nu_{R5}$ and $\tilde\nu_{R6}$ are degenerate. In this limit, all SUSY loop diagrams are proportional to $(Y_\nu^* M_R^*\, Y_\nu^\dagger)_{aa'}=2K_{aa'}(M_5^2-M_6^2)$. Then, if both neutrinos and sneutrinos are independently degenerate, one should expect all loop corrections to vanish, and thus no fine-tuning on light neutrino masses. We attribute this, on the one hand, to the cancellation of additional LNV terms on the neutrino side, and on the other hand, to the fact that in the unbroken SUSY limit a mass degeneracy for neutrinos would also imply degenerate sneutrino masses.

If $\Delta M_{65}\neq0$, then both SUSY and non-SUSY contributions can be large, so we need them to interfere destructively. As a first step, we have confirmed the result of~\cite{CandiadaSilva:2020hxj} in their SUSY-conserving limit, where $M_{\rm SUSY},\,\mu\to0$ and $\tan\beta\to1$. Here, the only non-vanishing SUSY contribution is the gaugino-higgsino $(\delta M^{\rm irr}_L)^{gh}$, which precisely cancels $(\delta M_L)^{\rm 2HDM}$.

However, once one considers broken SUSY, the cancellations become inefficient, with $(\delta M^{\rm irr}_L)^{gh}$ being subdominant with respect to $(\delta M_L)^{\rm 2HDM}$. We find destructive interference to be more likely if $m_{\tilde L}$, $m_{\tilde\nu_R}$, $M_{1,2}$ and $\mu$ are relatively light, or if heavy neutrino masses are close to the SUSY scale. However, due to the lack of experimental evidence in favour of SUSY, the sparticles must be heavy, making it very difficult to have large cancellations involving $(\delta M^{\rm irr}_L)^{gh}$ in the region accessible to heavy neutrino searches.

An important point to consider is that, once SUSY is broken, the pure gaugino $(\delta M^{\rm irr}_L)^{gg}$ can play an important role. For $a_\nu=0$, we find that the overall sign of this contribution depends on the relative sign between $\mu$ and $M_{1,2}$. In the following, we choose $\mu<0$ and $M_{1,2}>0$, which guarantees destructive interference with $(\delta M_L)^{\rm 2HDM}$. In this case, a light SUSY spectrum is again favoured, but can now be enhanced by large $|\mu|$. This is also the case when $a_\nu$ is large and positive.

\begin{figure}[tp]
\includegraphics[width=0.49\textwidth]{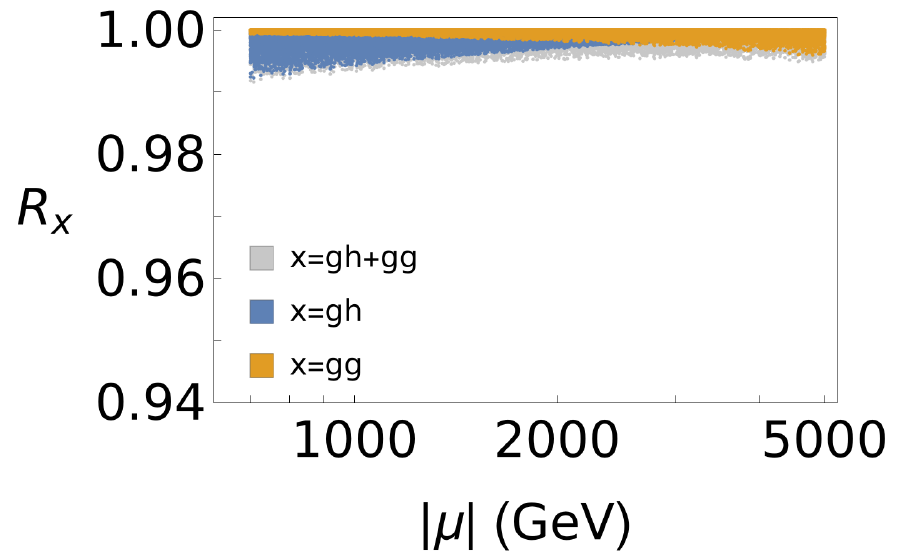}\hfill
\includegraphics[width=0.49\textwidth]{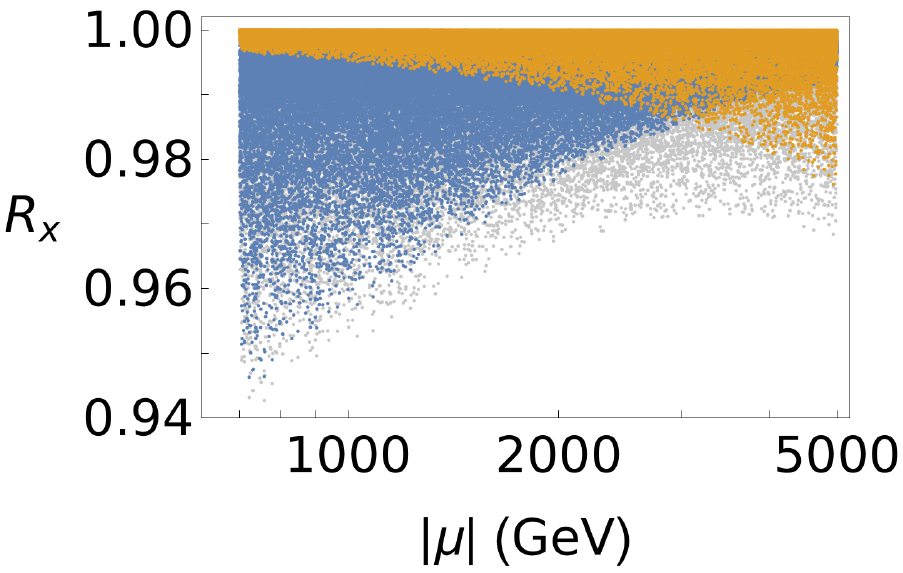}
\caption{$R_x$ parameter for increasing $|\mu|$, with $M_5=40$ ($200$) GeV on the left (right). Gaugino-higgsino (pure gaugino) cancellations are shown in blue (orange), with combined contribution in grey.}
\label{Fig:ScatterDegenerate}
\end{figure}
In order to illustrate our findings, we define $R_x=1+[(\delta M^{\rm irr}_L)^{x}/(\delta M_L)^{\rm 2HDM}]$ as a measure of the amount of cancellation possible between the SUSY and non-SUSY loop corrections, with $x=gh+gg,\,gh,\,gg$. Notice that the flavour structure of $(\delta M^{\rm irr}_L)^{x}$ and $(\delta M_L)^{\rm 2HDM}$ effectively cancels, leaving $R_x$ without flavour indices.

A scatter plot of $R_x$ as a function of $|\mu|$ is shown in Figure~\ref{Fig:ScatterDegenerate}, for two values of heavy neutrino mass, $M_5$. In the scan we have varied $-\mu,M_2,\,\,m_{\tilde L},\,m_{\tilde \nu}$ logarithmically between 700 and 5000~GeV, but allowing a soft mass $m_{\tilde\nu}$ as low as 0.1~GeV. We have set $M_1=M_2/2$ and $a_\nu=0$. Results are not strongly sensitive to $\Delta M_{65}$, nor any other parameter. In the Figure we confirm that cancellations are driven by $(\delta M^{\rm irr}_L)^{gh}$ for small $|\mu|$, and taken over by $(\delta M^{\rm irr}_L)^{gg}$ as $|\mu|$ grows. The combined contributions are relevant for intermediate values of $|\mu|$. Here, we confirm that the SUSY contribution cannot lead to strong cancellations any more, being less than $1\%$ ($5\%$) for $M_h=40$ (200) GeV, and thus cannot solve the fine-tuning problem.

\begin{figure}[tp]
\includegraphics[width=0.49\textwidth]{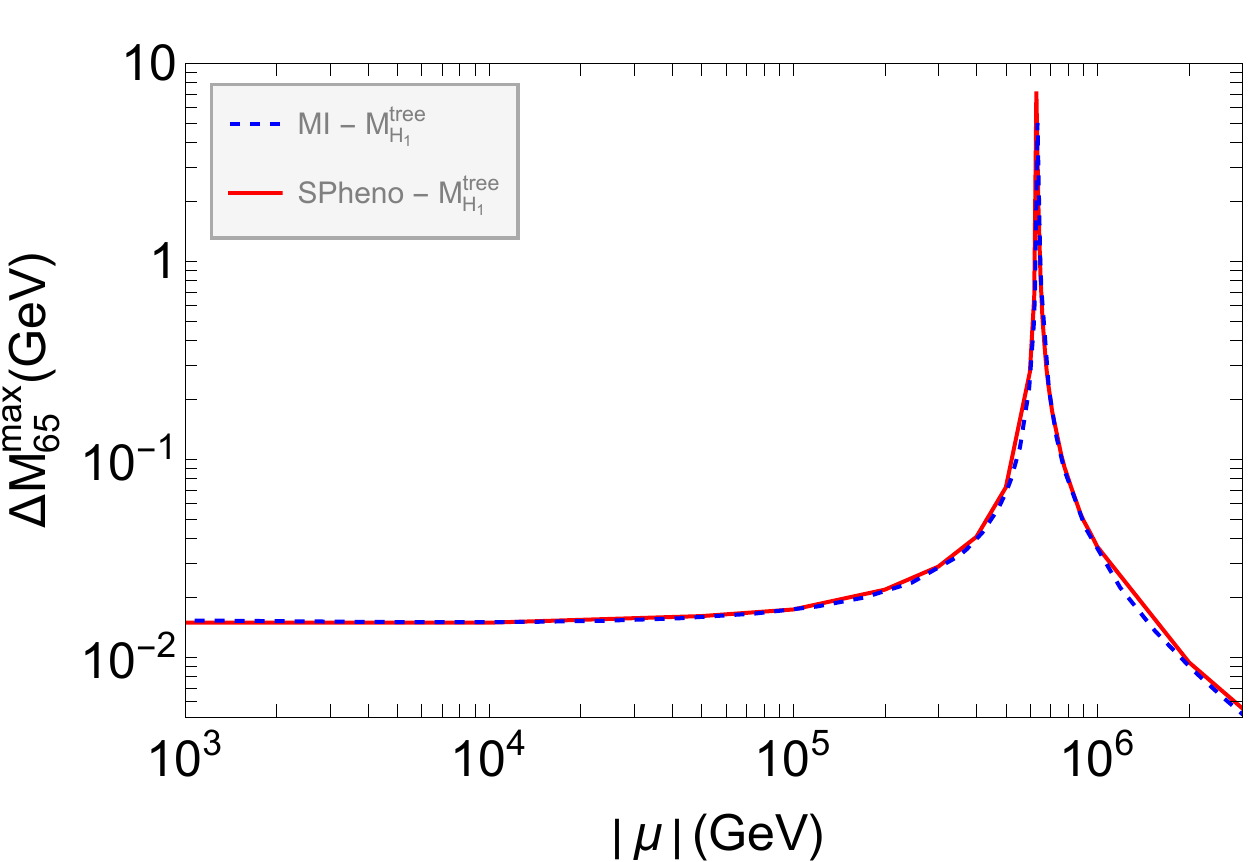}\hfill
\includegraphics[width=0.49\textwidth]{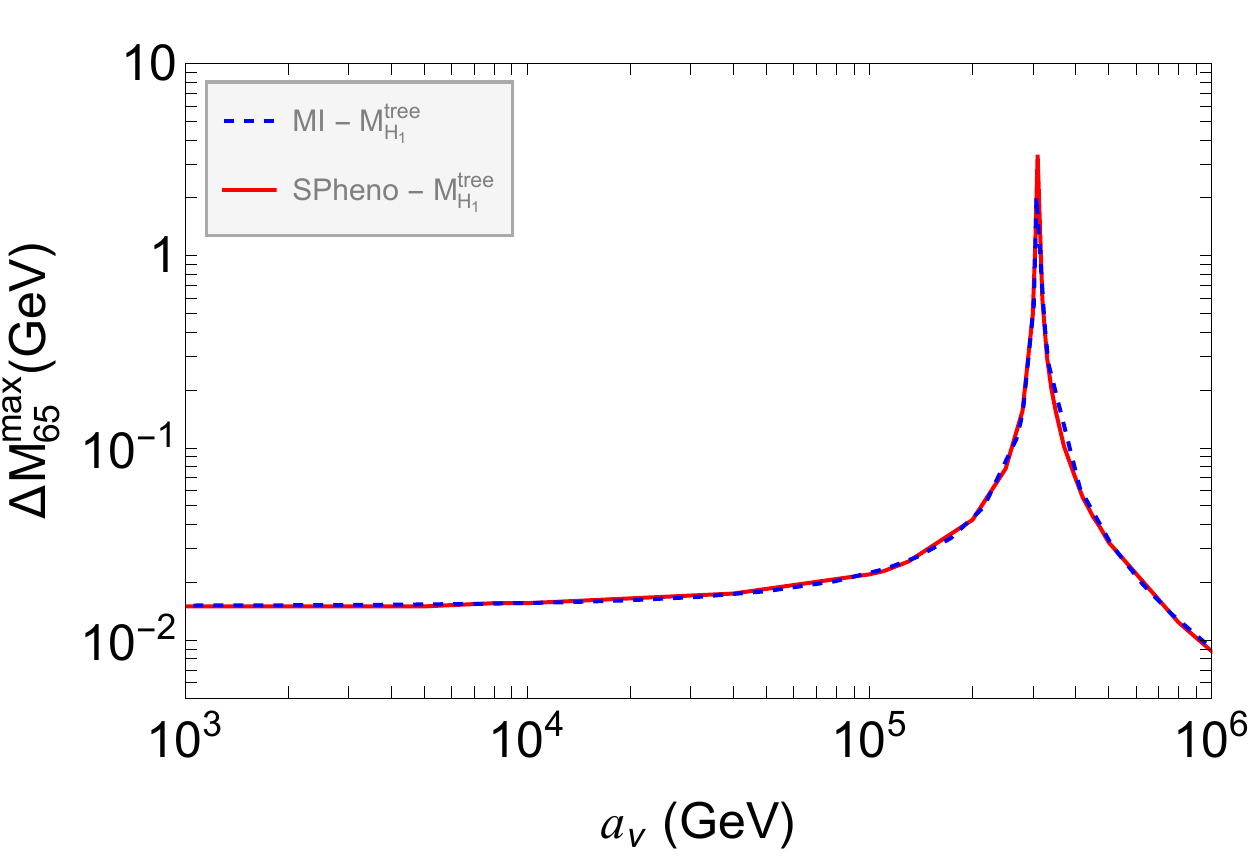}
\caption{Maximum allowed $\Delta M_{65}$ as a function of $|\mu|$ ($a_\nu$) on the left (right), for the SUSY spectrum considered in~\cite{CandiadaSilva:2020hxj}, with $\tan\beta=2$. We set $\gamma_{56}=8$, such that $|U_{\mu5}|^2=1.3\times10^{-7}$.}
\label{Fig:MaxDeltaMDegSUSYred}
\end{figure}

It is then of interest to understand the results of~\cite{CandiadaSilva:2020hxj}, who are able to relax constraints on LNV within their radiative inverse seesaw. Even though their setup is somewhat different from ours, the cancellation mechanisms are the same, and should be comparable regardless of the exact scenario in use. In order to shine light on the matter, we plot in Figure~\ref{Fig:MaxDeltaMDegSUSYred} the maximum $\Delta M_{65}$ allowed by requiring the full (SUSY + non-SUSY) loop corrections to be less than $50\%$, as a function of $|\mu|$ or $a_\nu$. The rest of the parameters are set as in Table II in~\cite{CandiadaSilva:2020hxj}. Within the Figure, we show curves obtained with our mass-insertion formulae, compared with those from \texttt{SPheno}~\cite{Porod:2003um,Porod:2011nf}, which performs the exact calculation using the tree-level Higgs mass\footnote{The model was implemented using \texttt{SARAH}~\cite{Staub:2008uz,Staub:2009bi,Staub:2010jh,Staub:2012pb,Staub:2013tta,Staub:2015kfa}}.

The first thing we notice is that, as verified by \texttt{SPheno}, the mass-insertion approximation holds very well, even for extremely large $|\mu|$ and $a_\nu$. In addition, we also find that there exist values for both parameters where SUSY and non-SUSY contributions cancel, allowing for a very large $\Delta M_{65}$. These are in the ballpark of the corresponding values reported in~\cite{CandiadaSilva:2020hxj}. The explanation for this is that $(\delta M^{\rm irr}_L)^{gg}$ has become very large, enhanced by either $|\mu|$ or $a_\nu$, and can cancel $(\delta M_L)^{\rm 2HDM}$, thus relaxing the LNV constraint on $\Delta M_{65}$. In fact, we see that the maximum $\Delta M_{65}$ decreases considerably after this cancellation, meaning that from this point $(\delta M^{\rm irr}_L)^{gg}$ not only cancels but exceeds $(\delta M_L)^{\rm 2HDM}$, needing an even smaller $\Delta M_{65}$ to be under control. 

What we conclude is that, in order to achieve the required cancellation, it is essential to select very precise values for either $|\mu|$ or $a_\nu$, greatly enhancing the pure gaugino contribution. Unfortunately, comparing this result with Figure~\ref{fig:loopfunction}, it can be argued that in this scenario the fine-tuning of the neutrino sector has been transferred to the SUSY sector, although this time without a symmetry such as LN to justify it.

Apart from this issue, we believe the large $|\mu|,\,a_\nu$ solutions have additional problems, which need to be addressed. Regarding $|\mu|$, the SUSY minimisation conditions would lead to a second situation with large fine-tuning, as the soft Higgs masses would need to have very special values to trigger electroweak symmetry breaking and reproduce the observed $Z$ mass simultaneously. It is likely this would also convey very large loop corrections to the light Higgs mass, leading to a third fine-tuning. On the other hand, for $a_\nu$, it was argued in~\cite{Faber:2019mti} that in order to avoid charge-breaking minima, one had to satisfy:
\begin{equation}
(a_\nu+M_R)^2\leq3(m^2_{H_u}+|\mu|^2+m_{\tilde L}^2+m_{\tilde\nu_R}^2+M_R^2+B_\nu)~,
\end{equation}
which is unlikely to hold given the benchmark spectrum. Thus, we do not consider the cancellations featured in~\cite{CandiadaSilva:2020hxj} to be a generic feature of the $\nu_R$MSSM, but to rather require additional ingredients beyond the simple structure of the model.

There does exist an alternative way of slightly improving the cancellations which, although inelegant\footnote{It is very unlikely that a high-scale model would provide such a spectrum after running the RGEs. Also, even in the case of a moderate splitting, the RGEs would generate off-diagonal soft terms, likely leading to problems with lepton flavour violating processes.}, does not require such large parameters. A direct inspection of Eqs.~(\ref{eq:gaugino_higgsino2}) and~(\ref{eq:gaugino_gaugino2}) shows that the $\tilde\nu_{R5}$ and $\tilde\nu_{R6}$ terms have opposite signs, meaning that the SUSY contribution is diminished when R-sneutrinos are degenerate. Then, if R-sneutrino masses are different, and if one of these is very large such that the corresponding R-sneutrino is decoupled, the SUSY contribution is maximised. However, the choice of which R-sneutrino needs to be decoupled depends on the neutrino sector. For example, we find that if $M_6>M_5$, then it is $\tilde\nu_{R5}$ who must be decoupled in order to guarantee destructive interference (with $\mu<0$ as before). In other words, the R-sneutrino hierarchy needs to be inverted with respect to the one for heavy neutrinos. 

\begin{figure}[tp]
\includegraphics[width=0.49\textwidth]{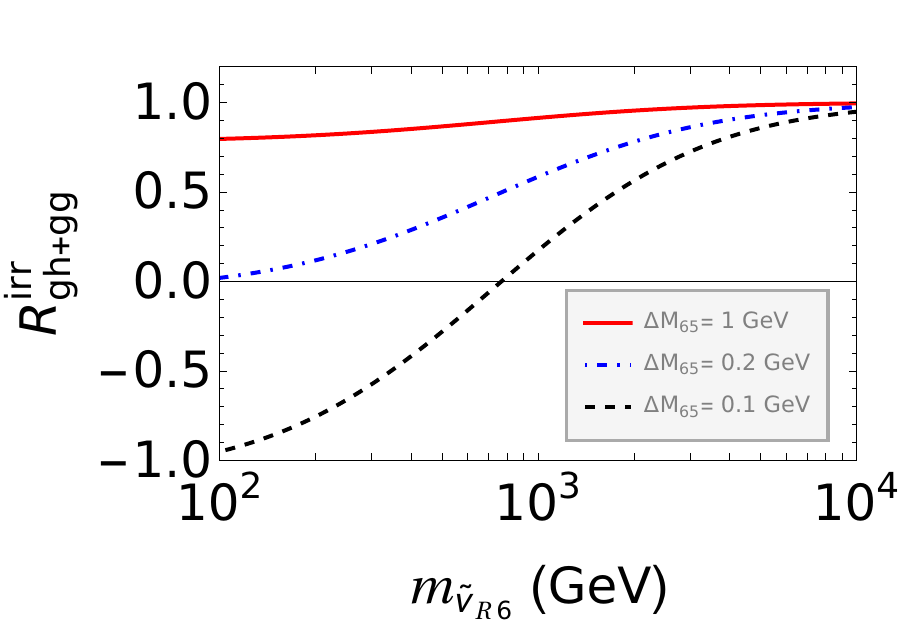}\hfill
\includegraphics[width=0.49\textwidth]{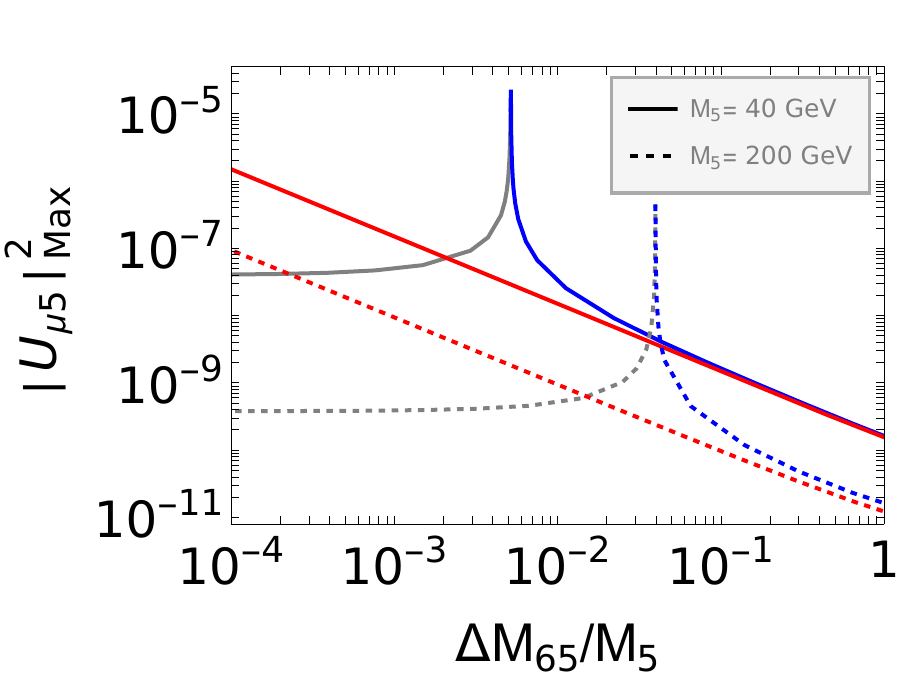}
\caption{Left: Dependence of $R_{gh+gg}$ with $m_{\tilde\nu_{R6}}$, for several values of $\Delta M_{65}$. Heavy neutrino mass is set to $M_5=40$~GeV. Right: Maximum allowed value of $|U_{\mu5}|^2$, as a function of $\Delta M_{65}/M_5$, for two values of $M_5$. Limits on for the Standard Seesaw are shown in red. Blue and grey lines show limits for the spectrum described in the text. For grey lines, the SUSY contribution is larger the the one from 2HDM.}
\label{Fig:NonDegenerateIrred}
\end{figure}
Given the properties of the loop functions, the most effective hierarchy leading to cancellations is $m_{\tilde\nu_{R6}}\ll m_{\tilde L}<-\mu,\,M_1,\,M_2\ll m_{\tilde\nu_{R5}}$. This suggests a spectrum similar to that of~\cite{Cerna-Velazco:2017cmn,Masias:2021uga}, which avoided LHC constraints and provided an R-sneutrino dark matter candidate~\cite{Faber:2019mti}. Considering this hierarchy, we show $R_{gh+gg}$ as a function of $m_{\tilde\nu_{R6}}$ on the left panel of Figure~\ref{Fig:NonDegenerateIrred}, for different neutrino mass splittings. The model parameters are $m_A=1.1$~TeV, $\tan\beta=8$, $m_{\tilde L}=500$~GeV, $M_2=750$~GeV, $M_1=M_2/2$, $\mu=-1$~TeV, and $a_\nu=0$. The Figure shows that, for a fixed $\Delta M_{65}$, the value of $R_{gh+gg}$ increases with $m_{\tilde\nu_{R6}}$. Thus, cancellations are stronger for $m_{\tilde\nu_{R6}}$ closest to the corresponding heavy neutrino mass (i.e.\ vanishing soft mass). Moreover, contrary to the degenerate case, the efficiency of the cancellation does depend on the heavy neutrino mass splitting, with smaller values of $R_{gh+gg}$ for smaller $\Delta M_{65}$. The reason is that the full SUSY contribution no longer depends on the neutrino splitting, meaning that reducing $\Delta M_{65}$ decreases only $(\delta M_L)^{\rm 2HDM}$, thus leading to lower $R_{gh+gg}$. Nevertheless, contrary to the degenerate case, this time the destructive interference can be substantial for moderate values of $|\mu|$, in some cases having the SUSY contribution exceeding the non-SUSY part ($R_{gh+gg}<0$).

The right panel of Figure~\ref{Fig:NonDegenerateIrred} compares the maximum $|U_{\mu5}|^2$ of the Standard Seesaw with that on our scenario, for the aforementioned spectrum. The bounds for the Standard Seesaw are shown in red, while the corresponding constraints for our model are shown in blue. Even though one can see a non-negligible relaxation of the bounds for small $\Delta M_{65}/M_5$, we find this effect vanishes when the splitting is large. Thus, we conclude that even for the non-degenerate case, SUSY does not relax the fine-tuning associated to heavy neutrinos with large mass splitting and mixing.

As a final comment, the Figure also shows grey curves for very small $\Delta M_{65}$, which place bounds much more stringent than those of the Standard Seesaw. Here, we find the SUSY contribution to be dominant, corresponding to negative $R_{gh+gg}$. To avoid these constraints one needs to take a heavier SUSY spectrum, in particular, larger $m_{\tilde\nu_{R6}}$ or $m_{\tilde L}$.

\subsection{Reducible Contributions}

Let us briefly comment on the three types of reducible contributions in our benchmark scenario, for degenerate sneutrinos and $a_\nu=0$, focusing on how to guarantee a cancellation with the non-SUSY part. First, we find that regardless of the spectrum, and for both signs of $\mu$, the gaugino-gaugino correction in Eq.~(\ref{eq:redgg}) gives destructive interference as long as $b_\nu<0$.

The gaugino-higgsino loop of Eq.~(\ref{eq:redgh}) also leads to cancellations for $b_\nu<0$, as long as $\mu<0$. If $\mu$ is positive, we find that the sign of the correction depends on the spectrum. However, we will not consider this possibility, as $\mu<0$ is also favoured by the irreducible gaugino-gaugino contribution.

As can be seen in in Eq.~(\ref{eq:redhh}), the higgsino-higgsino correction does not depend on $m_{\tilde L}$. Again concentrating on negative $\mu$, we find destructive interference for $b_\nu<0$ if $|\mu|$ is large. If $|\mu|$ is small, there exists a change in sign, requiring $b_\nu>0$ for cancellations. However, the latter possibility is in conflict with the other reducible corrections, which require $b_\nu<0$.

\begin{figure}[tp]
\includegraphics[width=0.49\textwidth]{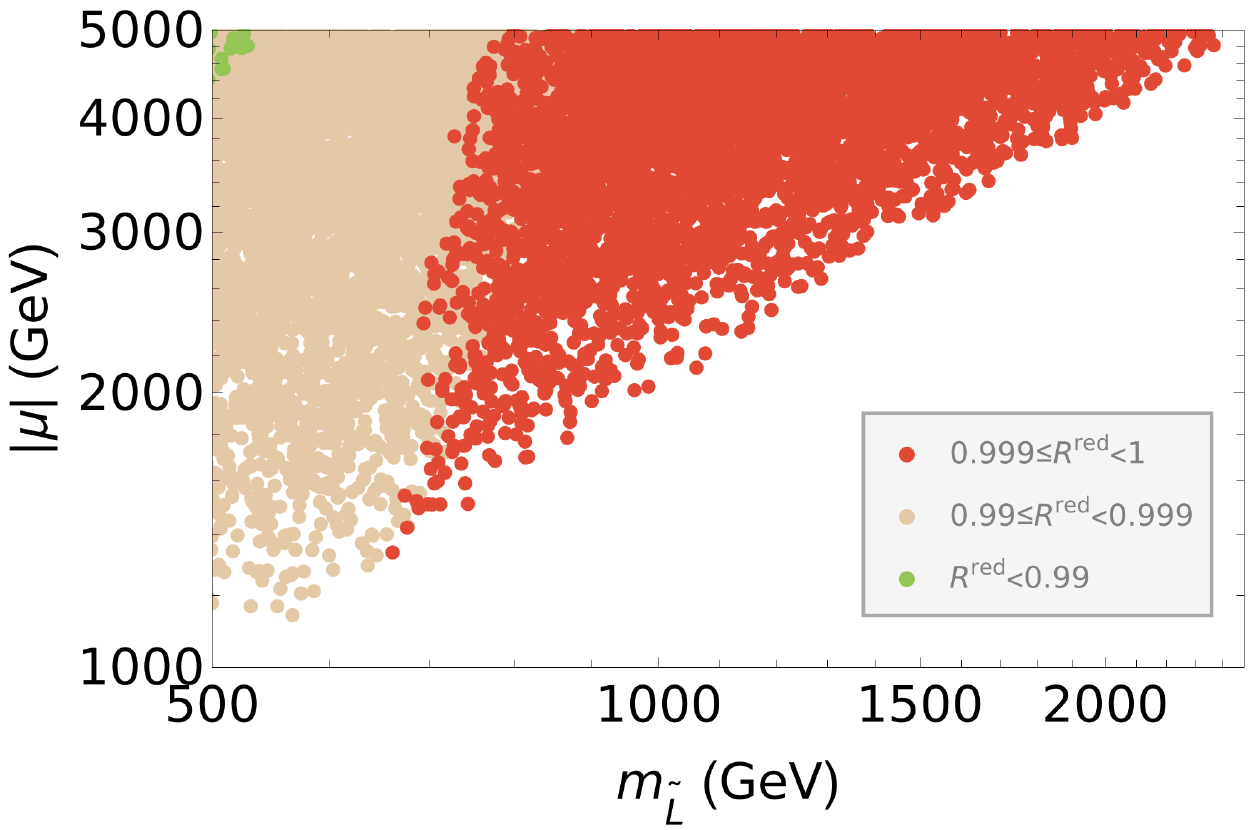}\hfill
\includegraphics[width=0.49\textwidth]{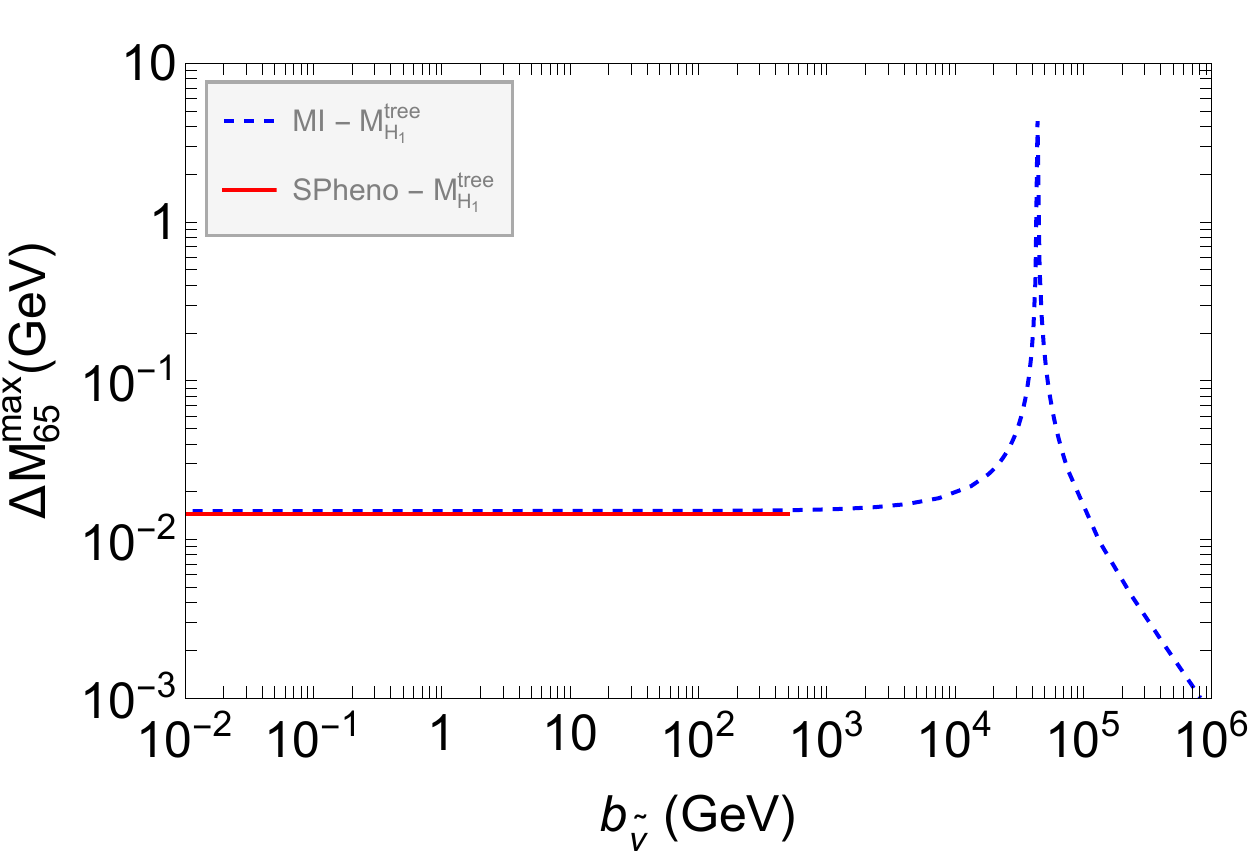}
\caption{Left: Values of $R^{\rm red}_{gg+gh+hh}$ parameter for different values of $m_{\tilde L}$ and $|\mu|$, with $M_5=200$~GeV. Right: Maximum allowed $\Delta M_{65}$ as a function of $b_\nu$, for the SUSY spectrum considered in~\cite{CandiadaSilva:2020hxj}, with $\tan\beta=2$. We set $\gamma_{56}=8$, such that $|U_{\mu5}|^2=1.3\times10^{-7}$.}
\label{Fig:ScatterRed}
\end{figure}
Thus, if we want all reducible and irreducible corrections to work together in cancelling the non-SUSY contribution, we need $\mu$ to be large and negative, as well as a negative $b_\nu$. In order to illustrate its behaviour we show, on the left panel of Figure~\ref{Fig:ScatterRed}, the corresponding $R^{\rm red}_{gg+gh+hh}$ considering only the reducible contribution, as a function of $|\mu|$ and $m_{\tilde L}$. Here, we have taken negative $\mu$, and varied $M_2=2M_1$ between $700$ and $5000$~GeV. Since effects are maximised for small $m_{\tilde\nu}$, we have set this parameter equal to $0.5$~GeV and, in order to avoid tachyonic states, set $b_\nu=M_5-5\,{\rm GeV}$. Results are shown only for $M_5=200$~GeV. 

In all points, we find that the gaugino-gaugino contribution dominates the correction, usually followed by gaugino-higgsino, and then higgsino-higgsino. We also find that the dependence on $\Delta M_{65}$ practically cancels with that of the non-SUSY part, so our results can be taken independent of the heavy neutrino mass splitting. However, within the evaluated parameter space, the cancellation is hardly above $1\%$, which happens for very large $|\mu|$ and very small $m_{\tilde L}$. The situation is worse for smaller $M_5$. When compared to the corresponding irreducible contributions, we find that the reducible part never rises above $0.6\%$ of the latter. Thus, we consider reducible contributions not worth considering any further.

Similarly to the irreducible case, the work in~\cite{CandiadaSilva:2020hxj} claims that one can find values of $b_\nu$ of $\ord{0.1~{\rm GeV}}$ that again cancel the non-SUSY loops. This time, we have not been able to reproduce their result. We show our attempt on the right panel of Figure~\ref{Fig:ScatterRed}, where we again plot the maximum $\Delta M_{65}$ in their benchmark scenario (note they use $\mu,\,b_\nu>0$). We do not find any cancellation around their expected value, coinciding with the prediction from~\texttt{SPheno}. Thus, it is possible that the $b_\nu$-based screening is a feature of models with radiative light neutrino masses. As a final remark, within the mass-insertion method, we again found that very large values of $b_\nu$ could be tuned in order to have the necessary destructive interference. These would work for $\mu>0$, and would be dominated by reducible higgsino-higgsino loops. However, when contrasted with~\texttt{SPheno}, we found that these would lead to tachyonic sneutrino states.

\section{Discussion}
\label{sec:Conclusions}

In this work we have briefly reviewed the problem of large loop corrections to light neutrino masses in the Type-I Seesaw model, which can be present in scenarios where the active-heavy mixing is large. It is well known that a good way of avoiding the problem is by assigning to the model a slightly broken LN symmetry, which forces the heavy neutrinos to appear as pseudo-Dirac states. This, however, can constrain LNV signals from appearing in collider searches.

We then evaluated a work appearing some years ago, which proposed considering a supersymmetric extension to the model as a way of keeping loop corrections under control. This study was motivated by the fact that in unbroken SUSY the quantum corrections to terms in the Superpotential are cancelled, leading to the hope than in the broken case a soft SUSY screening effect would follow. The expectation from this was that larger LNV parameters would be allowed on the neutrino sector, reflected on larger heavy neutrino mass splittings. This, in turn, would better motivate searches for LNV phenomena associated to a single heavy neutrino at colliders.

We thus performed a detailed analysis of the two irreducible and three reducible SUSY contributions to the loop corrected masses. We determined the regions of parameter space guaranteeing cancellations between the latter and the non-SUSY loops, concentrating on heavy neutrino mass ranges accessible to current collider experiments.

To summarise, we found the largest SUSY quantum corrections to be the irreducible ones. For the case of degenerate sneutrinos, with parameters under the TeV scale, we found no significant screening effect. We did corroborate that the pure gaugino loops could cancel the non-SUSY contributions for extremely large values of $|\mu|$ and $a_\nu$, but argued that doing so could cause problems in other sectors of the model. We also presented a non-elegant scenario with very non-degenerate sneutrinos, and found that the screening could be more efficient without needing too large $|\mu|$ or $a_\nu$. However, regardless of this, the relaxation of constraints on LNV was very mild.

It must be noted that none of the cases above, where the cancellations could be efficient, arises as a consequence of non-renormalisation theorems. The only SUSY screening contribution that does not rely on SUSY breaking, and thus could be attributed to the theorems, is the irreducible gaugino-higgsino loop which, as we have shown, only dominates for small $|\mu|$. None of the cases with efficient cancellations rely on this correction. Instead, they all need very specific values for the parameters, suggesting that what we are observing is a transfer of fine-tuning from the neutrino sector to the SUSY sector of the model. In our opinion, even though SUSY screening does sound appealing in principle, in practice it does not seem reasonable to bring in the whole supersymmetric framework to address this issue.

\section*{Acknowledgements}

The authors would like to thank Werner Porod for thorough discussions and a careful reading of the draft. We also would like to thanks Apostolos Pilaftsis for clarifications regarding their work. We acknowledge funding by the {\it Direcci\'on de Gesti\'on de la Investigaci\'on} at PUCP, through grant DGI-2021-C-0020, and have been also been supported by the DAAD-CONCYTEC project 131-2017-FONDECYT. O.S.N.\ was funded by grant 236-2015-FONDECYT. 

\bibliographystyle{unsrt}
\bibliography{bibl}
\end{document}